\documentclass[review]{elsarticle}

\usepackage{lineno,hyperref,amsmath}
\usepackage{wasysym,amssymb,gensymb}

\modulolinenumbers[5]

\journal{Journal of \LaTeX\ Templates}

\begin{document}

\begin{frontmatter}

\title{A Reassessment of the Quasi-Simultaneous Arrival Effect in Secondary Ion Mass Spectrometry}


\author{Ryan Ogliore}
\address{Department of Physics, Washington University in St. Louis, St. Louis, MO 63130, USA}

\author{Kazuhide Nagashima, Gary Huss}
\address{Hawai`i Institute of Geophysics and Planetology, University of Hawai`i at M\={a}noa, Honolulu, HI 96822, USA}

\author{Pierre Haenecour}
\address{Lunar \& Planetary Laboratory, University of Arizona, Tucson, AZ 85721, USA}

\begin{abstract}
Quasi-simultaneous arrival (QSA) effects in secondary ion mass spectrometry can create mass-indepedent inaccuracies in isotope measurements when using electron multiplier detectors (EMs). The simple Poisson statistical model of QSA does not explain most experimental data. We present pulse-height distributions (PHDs) and time-series measurements to better study QSA. Our data show that PHDs and the distribution of multiple arrivals on the EM are not consistent with the Poisson model. Multiple arrivals are over-dispersed compared to Poisson and are closer to a negative binomial distribution. Through an emission-transmission-detection model we show that the QSA correction depends on the non-Poisson emission of multiple secondary ions, the secondary ion energy distribution, and other factors, making an analytical correction impractical. A standards-based correction for QSA is the best approach, and we show the proper way to calculate standards-normalized $\delta$ values to minimize the effect of QSA.

\end{abstract}


\end{frontmatter}


\section{Introduction}
Measurements of the isotopic composition of micrometer-sized and smaller grains \textit{in-situ} by secondary-ion mass spectrometry (SIMS) has revolutionized cosmochemistry, geochemistry, biology, and other fields. Modern SIMS instruments employ two types of detectors: current- (or integrated charge) measuring Faraday cups and ion-counting electron multipliers. To achieve the smallest primary beam, and therefore, the best spatial resolution, it is necessary to decrease the primary beam current to less than one nanoamp. Secondary ion currents for these measurement conditions typically are small compared to the Johnson noise of the Faraday cup preamplifier, and so must be measured by electron multipliers (EMs). Additionally, secondary-ion raster imaging requires fast collection and processing of the secondary ion signal, which is possible with electron multipliers and not possible with Faraday cups. The accuracy of a measurement is affected by myriad factors such as deadtime correction, variable electron multiplier gain, and the quasi-simultaneous arrival (QSA) effect \cite{slodzian2001precise}. In this paper we present a reassessment of the QSA effect in SIMS with the goal of improving accuracy for measurements of small samples, such as cometary material returned by NASA's Stardust mission \cite{ogliore2015oxygen}, and in the near future, asteroid regolith samples returned by NASA's OSIRIS-REx and JAXA's Hayabusa2 missions.

\subsection{The Poisson Model of QSA}

A simplified model of the QSA effect is described in \cite{slodzian2004qsa}. In this model, primary ions impact the sample surface. The number of secondary ions created by each primary ion impact is modeled as a Poisson process with mean equal to $K$. The value of $K$ can be determined by the secondary ion count rate divided by the primary ion count rate, the secondary-to-primary ratio. The transmission of secondary ions from the sample surface to the detector is ignored in this simplified model---quasi-simultaneous arrival results directly from quasi-simultaneous emission.

The true number of emitted secondary ions per primary ion impact, $N_{true}$, is the expectation value of the secondary ion emission process, assumed to be Poisson: 
\begin{equation}
    N_{true}=\sum_{n=1}^{\infty} (n) \mathcal{P}(K,x=n) = \sum_{n=1}^{\infty} (n) \frac{K^n e^{-K}}{n!}=K
    \label{equation:poiss}
\end{equation}
After emission from the sample, the secondary ions will be accelerated, travel through the mass spectrometer, and hit the first dynode of the EM. Electrons created by the secondary ion impact are accelerated and impact the next EM dynode, where more electrons are created. A series of dynodes creates an electron cascade that is amplified after the last dynode by a pre-amp. The signal from the pre-amp is processed by a discriminator, then these pulses are digitized as an output count rate.

Pulses above the user-defined threshold voltage will be counted by the discriminator and trigger the user-defined deadtime. If a second pulse arrives within this deadtime window, it will not be counted. If the arrival time difference is comparable to the width of the pulse at the output of the preamp ($\sim$10~ns), or smaller, the two pulses will appear as one at the preamp output, with a height equal to the convolution of the two individual pulses.  

These ``quasi-simultaneous'' ions will be measured as a single event by the pulse-counting electron-multiplier detector.  The measured counts per primary ion impact $N_{measured}$ can be calculated similar to Equation \ref{equation:poiss}, except one count is measured for all multiple emissions:

\begin{equation}
    N_{measured}=\sum_{n=1}^{\infty} (1) \mathcal{P}(K,x=n)=1-\mathcal{P}(K,x=0)=1-e^{-K}
\end{equation}

The ratio $N_{true}/N_{measured}$ is then:

\begin{equation}
    \label{eqn:poissonexactcorr}
    \frac{N_{true}}{N_{measured}}=\frac{K}{1-e^{-K}}
\end{equation}

This is an easy quantity to calculate and does not need to be simplified further. In the literature however, this quantity is simplified using a series expansion in $K$, the secondary to primary count rate ratio. However, $K$ can reach values up to $\sim$1, making the higher order terms significant (up to 5\%), so this simplification seems both unnecessary and inaccurate. Nonetheless, the series expansion yields:

\begin{equation}
    \frac{N_{true}}{N_{measured}}=\frac{K}{1-e^{-K}} = 1+\frac{K}{2}+\frac{K^2}{12}-\frac{K^4}{720} + \mathcal{O}(K^5)
\end{equation}

The first two terms, $1+K/2$, are typically retained and higher order terms are dropped \cite{slodzian2004qsa}. 

The justification for the above series expansion is that while SIMS isotope ratios (with the more abundant isotope in denominator) are often measured to be lower with larger $K$, the effect does not follow the predictions of this model \cite[][and references therein]{jones2020statistical}. To account for this discrepency, instead of $1+K/2$, a correction of the form $1+\beta K$ is applied where $\beta$ is determined from measurements of standards. The values of $\beta$ vary widely---from 0.19 \cite{nishizawa2010micro} to 1.0 \cite{hillion2008effect}. A $\beta$ value not equal to $1/2$ invalidates the Poisson model described above. The $\beta \neq 1/2$ correction is not a modification of the above model, it is completely ad hoc and not derived from any physical or statistical model. QSA corrections can be very large (tens of per mil) compared to the desired precision of a SIMS measurements \cite[e.g., ][]{nguyen2017coordinated}. More importantly, however, QSA corrections are mass-independent because they affect the more abundant (higher secondary ion count rate)  isotope (typically in denominator of the isotope ratio) much more than the less abundant (numerator) isotope. Mass-independent effects are diagnostic of important processes in many fields such as cosmochemistry \cite{thiemens199612}, so instrumental mass-independent fractionation must be well understood. A physical model is needed to understand QSA and correct for it properly. 

\subsection{The Poisson Model of QSA}

Slodzian et al.\ \cite{slodzian2004qsa} proposed that the cause of the deviation of measured QSA effects from the above-described model may be due to ``inadequacy of Poisson statistics to describe the phenomenon or to other effects such as fractionations due to differences in ion selection generated by the change in $K$''. As mentioned above, the model of \cite{slodzian2004qsa} assumes that nothing in the mass spectrometer modifies the time distribution of secondary ions emitted from the sample (quasi-simultaneous emission equals quasi-simultaneous arrival). A more complete model that accurately predicts experimental data should take into account the true statistical nature of the quasi-simultaneous emission process, and the physics between emission and detection of the secondary ions at the EM.

To investigate these two phenomena, we present two different measurements in this paper: 1) high-precision measurements of pulse-height distributions using the Cameca NanoSIMS 50 at Washington University in St.\ Louis (Section \ref{pulseheightmethods}), and 2) time-series measurements of pulses measured at the output of the mono-collector EM preamp on the UH Cameca ims 1280 (Sections \ref{timeseriesmethods}).

\section{Pulse-Height Distributions}
\label{pulseheightmethods}
If two pulses arrive simultaneously at the first dynode of the electron multiplier, they combine to make a single pulse height, which would be twice as large as a single-count event on average. A measured histogram (a pulse-height distribution, or PHD) of these pulse heights would show QSA events as an excess of pulses in the high-voltage tail of the histogram. 

\subsection{Pulse-Height Distribution Model}

The probability for producing $n$ secondary electrons for one EM dynode is best modeled by the P\'{o}lya distribution \cite{wright1969detection}, a special case of the negative binomial distribution:

\begin{equation}
    P(n)=\frac{\mu^n}{n!}(1+b\mu)^{-n-1/b}\prod_{i=1}^{n-1}(1+bi)
\end{equation}

where $\mu$ is the average gain of each stage of the EM, and $\lambda$ and $b$ are non-negative constants. With $b=0$, the distribution is Poisson. For a sequence of $k$ dynodes:

\begin{equation}
    P_k(n)=\frac{\mu}{n} (P_k(0))^b \left(  \sum_{i=0}^{n-1} (n+ib-i) P_k(i) P_{k-1}(n-i)  \right)
\end{equation}

where $P_k(0)$ is the probability that the total number of secondaries is zero at the $k^{th}$ dynode, and is calculated:

\begin{equation}
    P_k(0)=
    \begin{cases}
      \left(1+b \mu(1-P_{k-1}(0))  \right)^{-1/b}, & \text{if}\ b>0 \\
      e^{\mu P_{k-1}(0)-\mu}, & \text{if}\ b=0
    \end{cases}
  \end{equation}
  
To start the calculation, we must specify the number of electrons per incident ion on the first dynode:

\begin{equation}
    P_0(n)=\delta(n=N_p)
\end{equation}

This is the Dirac delta function at $N_p$, the number of electrons per ion generated at the first dynode. All subsequent dynodes after the first are assumed to have the same mean number of electrons per incident electron: $N_e$. Therefore, the total gain of the EM with $k-1$ dynodes after the first dynode is: 
\begin{equation}
\mu=N_p (N_e)^{k-1}
\end{equation}

\subsection{Measurements of Pulse Height Distributions}

We collected pulse-height distributions of $^{28}$Si and $^{30}$Si on the Cameca NanoSIMS 50 at Washington University in St.\ Louis, using a Cs$^+$ primary beam and 10$\times$10~$\mu$m raster. We collected PHDs for $\sim$12~hrs at count rates of $\sim$250,000 cps by averaging together 500--600 individual scans. We varied the entrance slits to change $K$ from $<$0.01 to $\sim$0.1. High count rates on an EM for a long time can decrease the EM's efficiency, by decreasing $N_p$ and/or $N_e$. We checked the overall gain of the EM before and after the measurement by comparing the pulse-height distributions. We confirmed that the EM gain did not change significantly over the course of each individual measurement (Low $K$ and High $K$).

\begin{figure}
    \centering
    \includegraphics[width=0.45\textwidth]{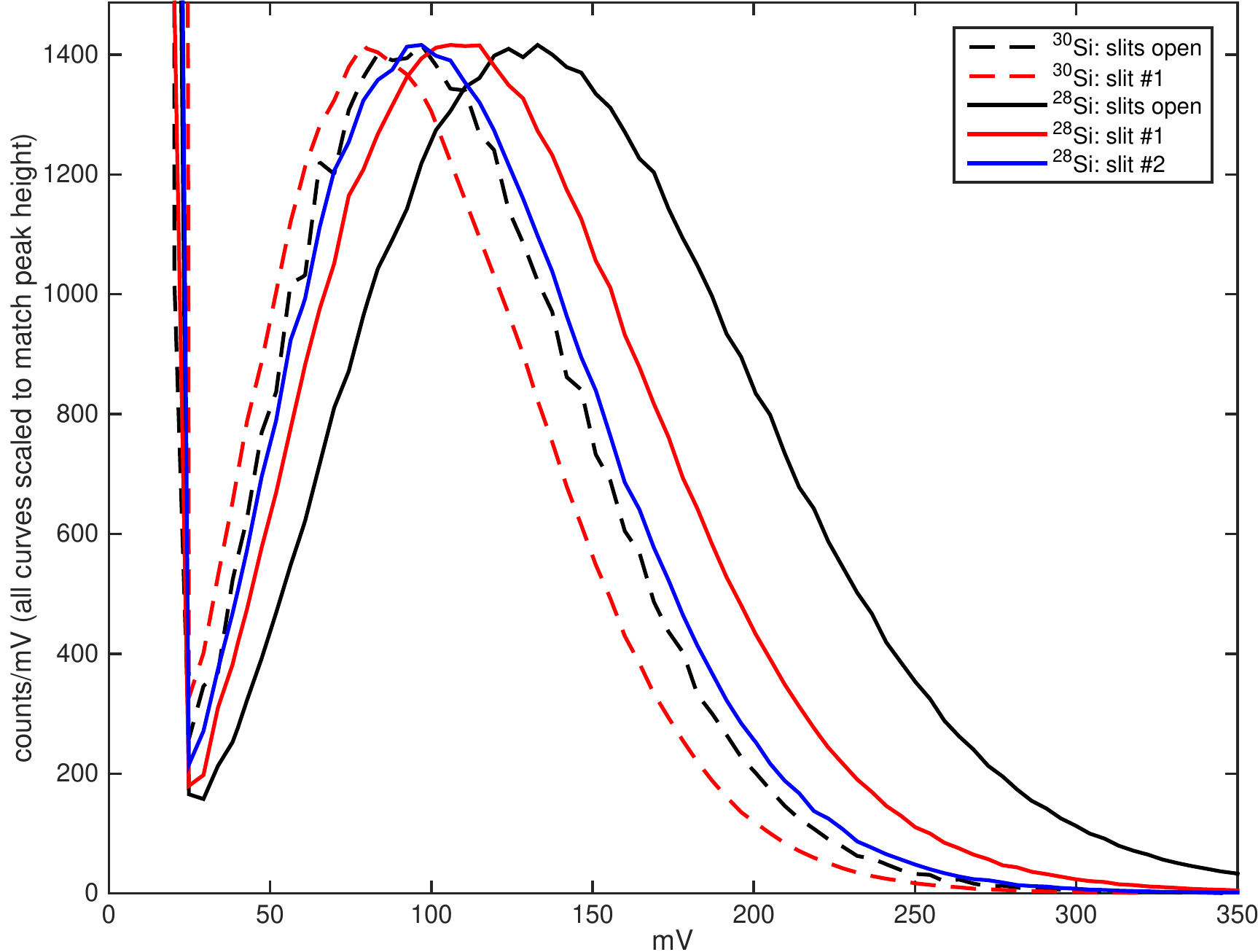}
    \includegraphics[width=0.45\textwidth]{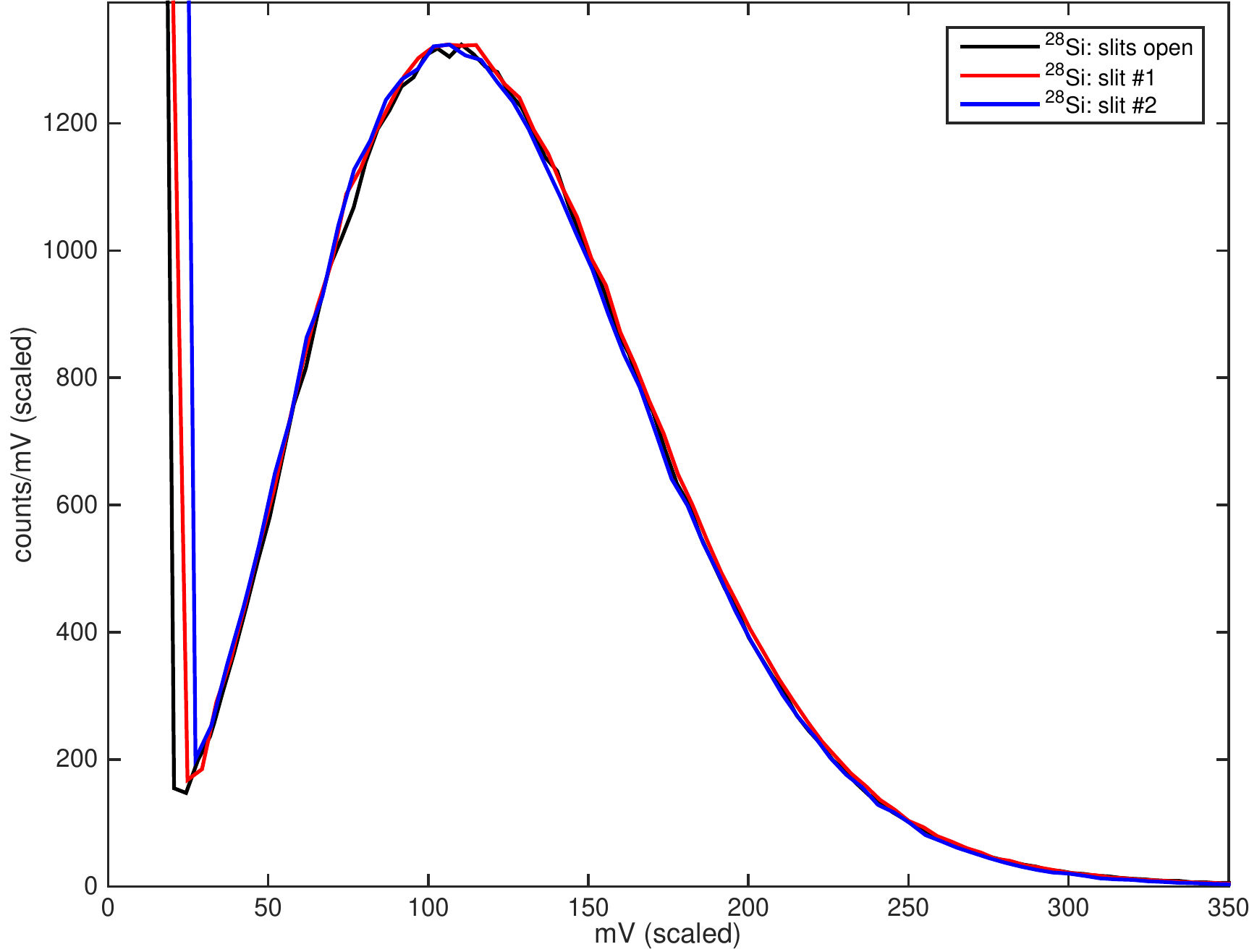}
    \caption{Left) Pulse-height distributions for entrance slits \#1, \#2, and the open slit for $^{28}$Si and $^{30}$Si. Right) Pulse-height distributions for $^{28}$Si with three different entrance slits (corresponding to $K \approx 0.01, 0.044, 0.083$), scaled to account for changes in $N_p$.}
    \label{fig:phd}
\end{figure}

The PHDs shown in the left panel of Figure \ref{fig:phd} have different characteristic widths and peak voltages. The widths and peak locations may change with species and the entrance slit. In the context of the above model, the cause of this difference is due to changes in $N_p$, the number of electrons created per incident ion on the first dynode of the EM. The first dynode of the EM is known to age, effectively decreasing $N_p$ with time. (To increase the lifetime of the EMs (which are expensive to replace), the EM bias voltage is periodically increased to offset the aging effect). Because the focused secondary ion beam impacts the detector in different places for different analyses, $N_p$ may also vary spatially. The location on the EM where the secondary ion beam hits is a sensitive function of instrument tuning (e.g., B field value, EM position, deflector values, etc.). Consequently, if the secondary ion beam strikes a slightly different place on the EM, due to changes in instrument tuning, $N_p$ will change. The right panel of Figure \ref{fig:phd} shows the three PHDs for $^{28}$Si with $K \approx 0.010, 0.044, 0.083$, with $N_p$ varied so that the distributions align.

\begin{figure}
    \centering
    \includegraphics[width=0.7\textwidth]{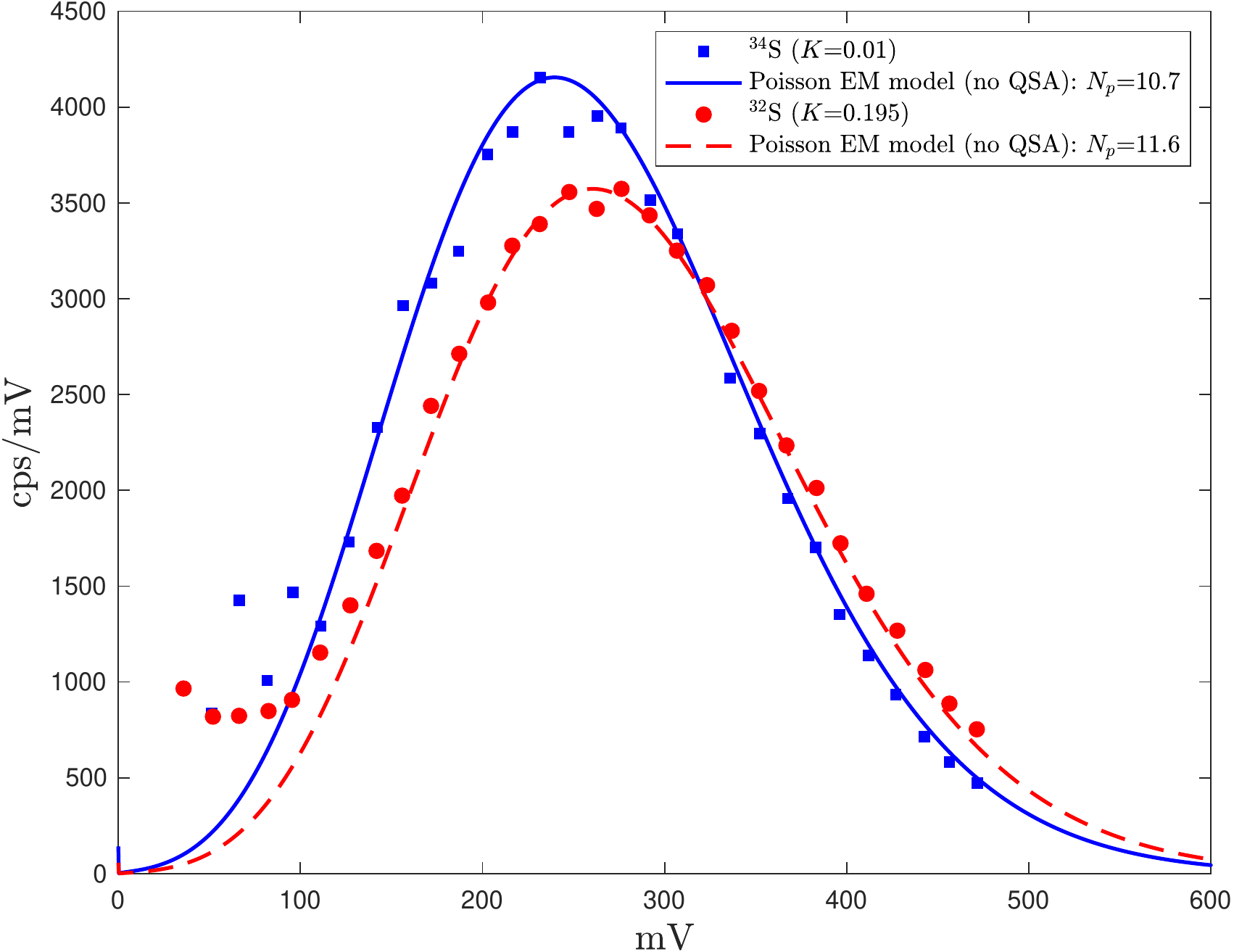}
    \caption{Pulse-height distributions from Figures 4 and 5 of \cite{slodzian2004qsa} for $^{32}$S (red circles) and $^{34}$S (blue squares). Blue solid line is a Poisson cascade model, without QSA events, of the electron multiplier with $N_p=10.7$ and $N_e=3$. Red dashed line is the same model with $N_p=11.6$.}
    \label{fig:phdslodzian}
\end{figure}

The change in pulse-height distributions reported by \cite{slodzian2004qsa} were interpreted as a QSA effect. We simulated these PHDs to investigate this claim. We employed a Poisson cascade model, as in \cite{slodzian2004qsa}, with five total dynodes (total computation time grows rapidly with the number of dynodes) with no multi-hit events due to QSA. Fixing $N_p=10.7$ and $N_e=3$, as in \cite{slodzian2004qsa}, we changed the total simulated gain to match the peak for $^{34}$S. (The technical document from the EM anufacturer, Hamamatsu photonics, says $N_p$ should be close to 20 for 10~keV impact energy. However, $N_p=20$ gives a distribution that is much narrower than the data.) Then we used these same conditions, still with no QSA effect, and changed only $N_p$ to match the PHD for $^{34}$S. With $N_p=11.6$ we were able to match the $^{32}$S data without QSA as well as the model by \cite{slodzian2004qsa} with QSA (Figure \ref{fig:phdslodzian}). We conclude that variable gain in the EM's first dynode between measurements of $^{34}$S and $^{32}$S by \cite{slodzian2004qsa} are more likely to be responsible for their different PHDs than the effect from QSA.


We investigated the effect of QSA in our measured PHDs with two $K$ (secondary to primary count rate ratio) values: $0.01$ and $0.083$. We use the $K=0.01$ PHD as a template (scaled so that both distributions have similar $N_p$), to which we will add QSA counts (see below) to fit the PHD for $K=0.083$.

To calculate the effect QSA has on the PHD, we assume the $K=0.01$ PHD ($f_{0}$) has no pulse contributions from QSA. To estimate the distribution of the $K=0.083$ PHD ($f_{H}$), we first calculate the PHD of two ions impacting the EM simultaneously so that the voltage pulses at the pre-amp output combine. We assume that the pulse-height distribution of these two pulses are identically distributed. As shown in Section \ref{measurementsofsecondaryionarrivals} and Figure \ref{fig:QSA_PHDs}, our time-series measurements of QSA pulses supports this assumption. The probability distribution of a pulse of height $z$ is given by the conditional probability:

\begin{equation}
    f_{double}(z)=\int_{-\infty}^{\infty} f_{0}(z)f_{0}(z-x)dx = (f_0 * f_0)(z)
\end{equation}

where $*$ is the convolution operator.

\begin{figure}
    \centering
    \includegraphics[width=0.5\textwidth]{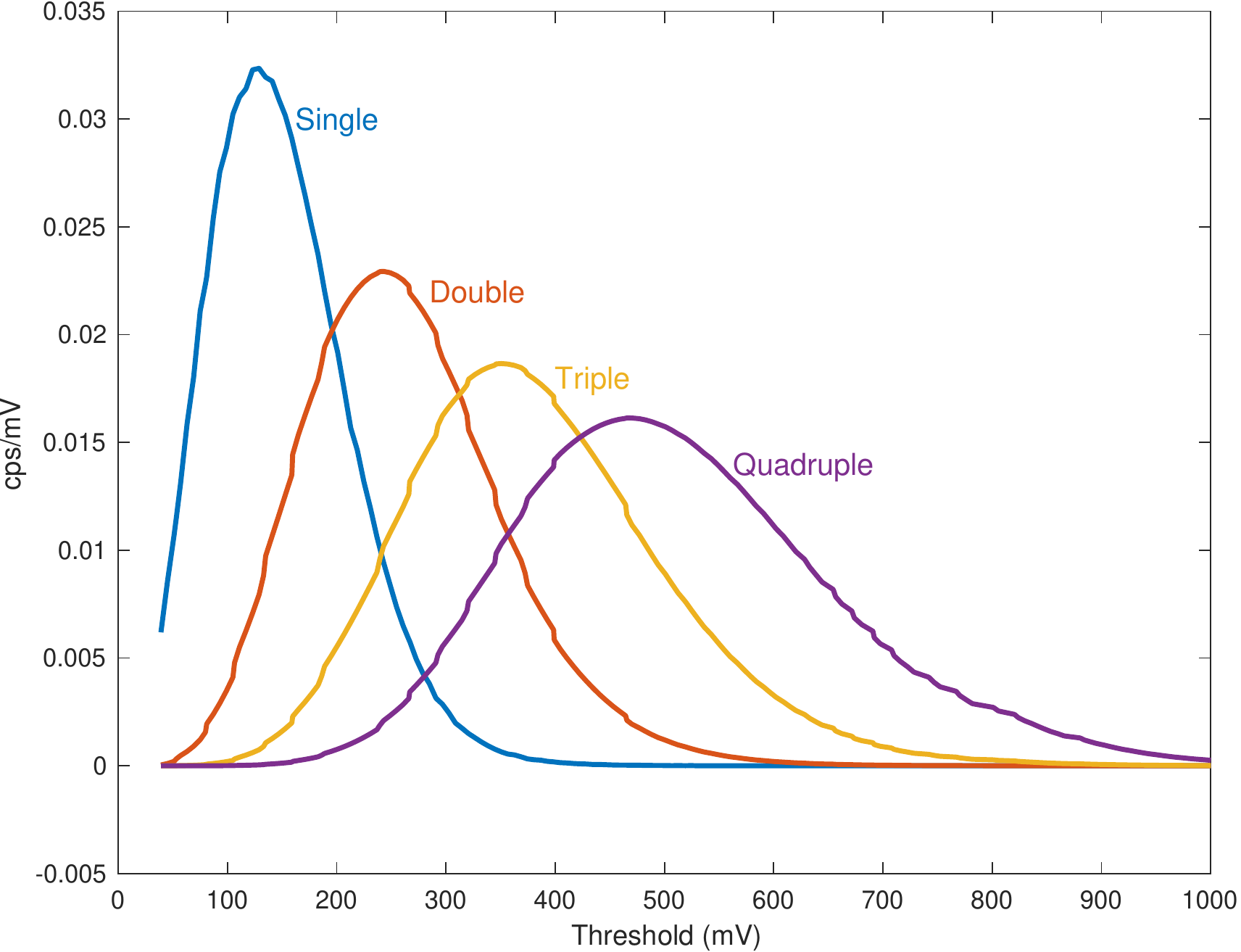}
    \caption{PHDs for simultaneous pulses at the first dynode of the EM, normalized to have the same area under the curve.}
    \label{fig:multipulses}
\end{figure}

Similarly, for three pulses arriving simultaneously, the probability distribution would be $(f_0 * f_0 * f_0)(z)$. These multi-hit PHDs are shown in Figure \ref{fig:multipulses}, which shows that the multi-hit pulses will substantially affect the high-voltage tail of the PHD.

The pulse-height distribution $f_{H}$ for a measurement with a secondary/primary ratio of $K$ can be modeled:

\begin{align}
    f_H=&(f_0)Poiss(1,K)+(f_0 * f_0)Poiss(2,K)\\ \nonumber
    &+(f_0 * f_0 *f_0)Poiss(3,K)+\cdots
\end{align}

We scale the $N_p$ of $f_0$ to match the peak of $f_H$. This is equivalent to simply multiplying the voltage value for each histogram bin in the pulse-height distribution by some factor. Since we averaged many individual pulse-height distributions, factors liek a decreasing secondary ion signal, or aging of the EM, will not affect the shape of our measured pulse-height distribution.

\subsection{QSA in Measured Pulse Height Distributions}

\begin{figure}
    \centering
    \includegraphics[width=0.5\textwidth]{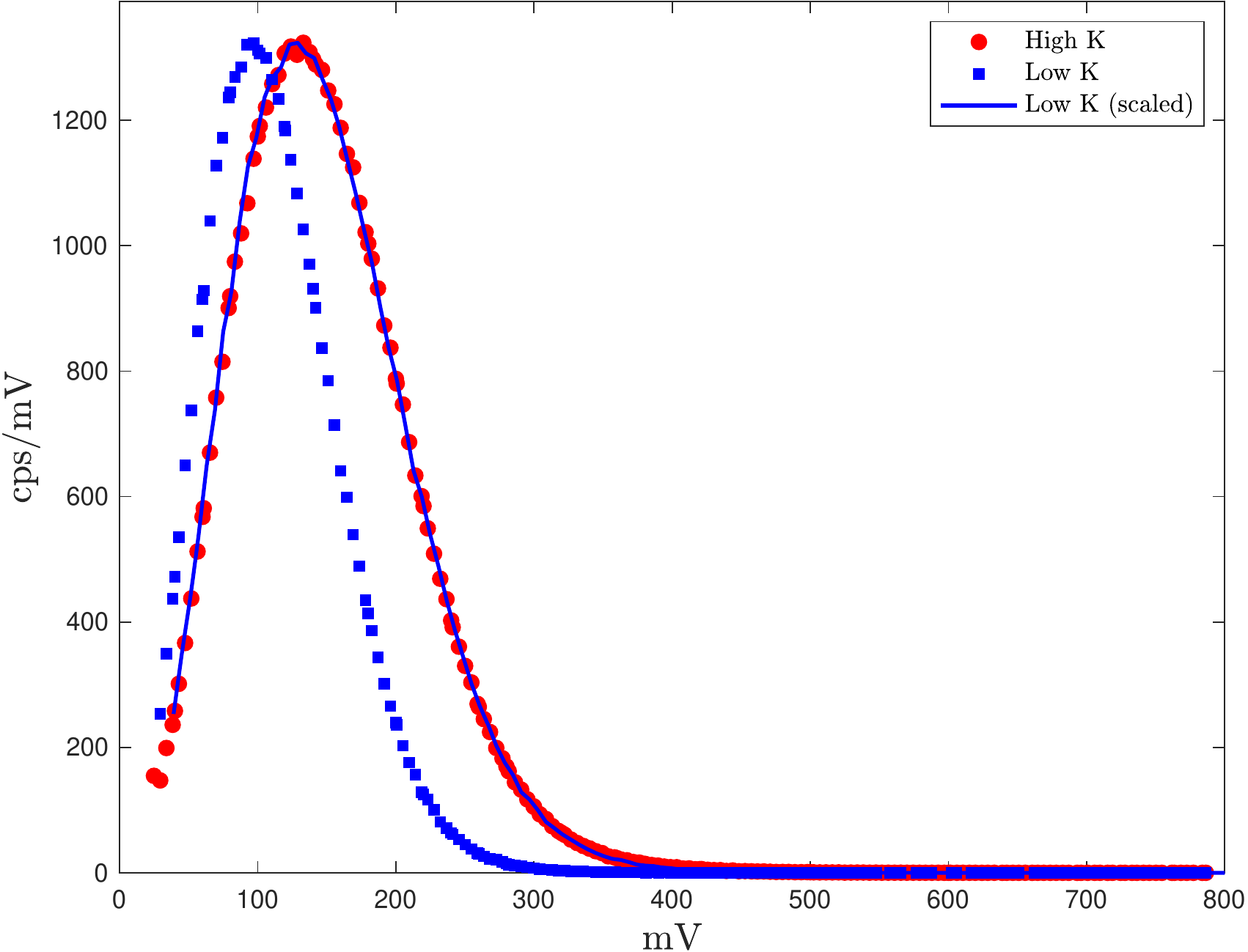}
    \caption{PHD for the low-$K$ measurement scaled to the high-$K$ measurement.}
    \label{fig:scaledphds}
\end{figure}

We focus on the high-voltage tails of the PHDs because this is where the QSA effect is most apparent. The data and model for $f_{0}$ and $f_{H}$ are shown in Figure \ref{fig:scaledphds}. We found that the $K=0.01$ scaled template (blue line) under-predicts the number of large pulses at the EM. This implies that there is some QSA contribution that is unaccounted for by the template. However, the $K=0.083$ QSA simulation (green line) greatly overestimates the number of high pulses. A $K$ value of $0.017$ (black line), five times smaller than the actual secondary to primary ion ratio, describes the PHD well.

We conclude from this analysis that the effect of QSA is significant and measurable, but does not follow the simple Poisson model with the measured secondary-to-primary ratio $K$ . In this case of $^{28}$Si measured on the NanoSIMS, the Poisson QSA model of \cite{slodzian2004qsa} overestimates the QSA effect.

A possible explanation is that because secondary ions are not arriving exactly simultaneously at the first EM dynode, this simplified model does not accurately model the effect on the PHD. If the QSA pulses are arriving at slightly different times, say 5~ns apart, the convolution of these two pulses will not be approximately equal to the sum, and the PHD will not change as we have modeled it. If there is a distribution of arrival times of the pulses, some close enough together so that the convolution of the pulses is approximately equal to the sum of the pulses, and some far enough apart so that this is not true, we will observe a QSA effect less than what is predicted by this model. In the following section, we show an example of three such pulses (Figure \ref{fig:closepeaks}), and show that differences in the arrival times of quasi-simultaneously emitted secondary ions is expected because of the energy distribution of the secondary ions.  

\begin{figure}
    \centering
    \includegraphics[width=\textwidth]{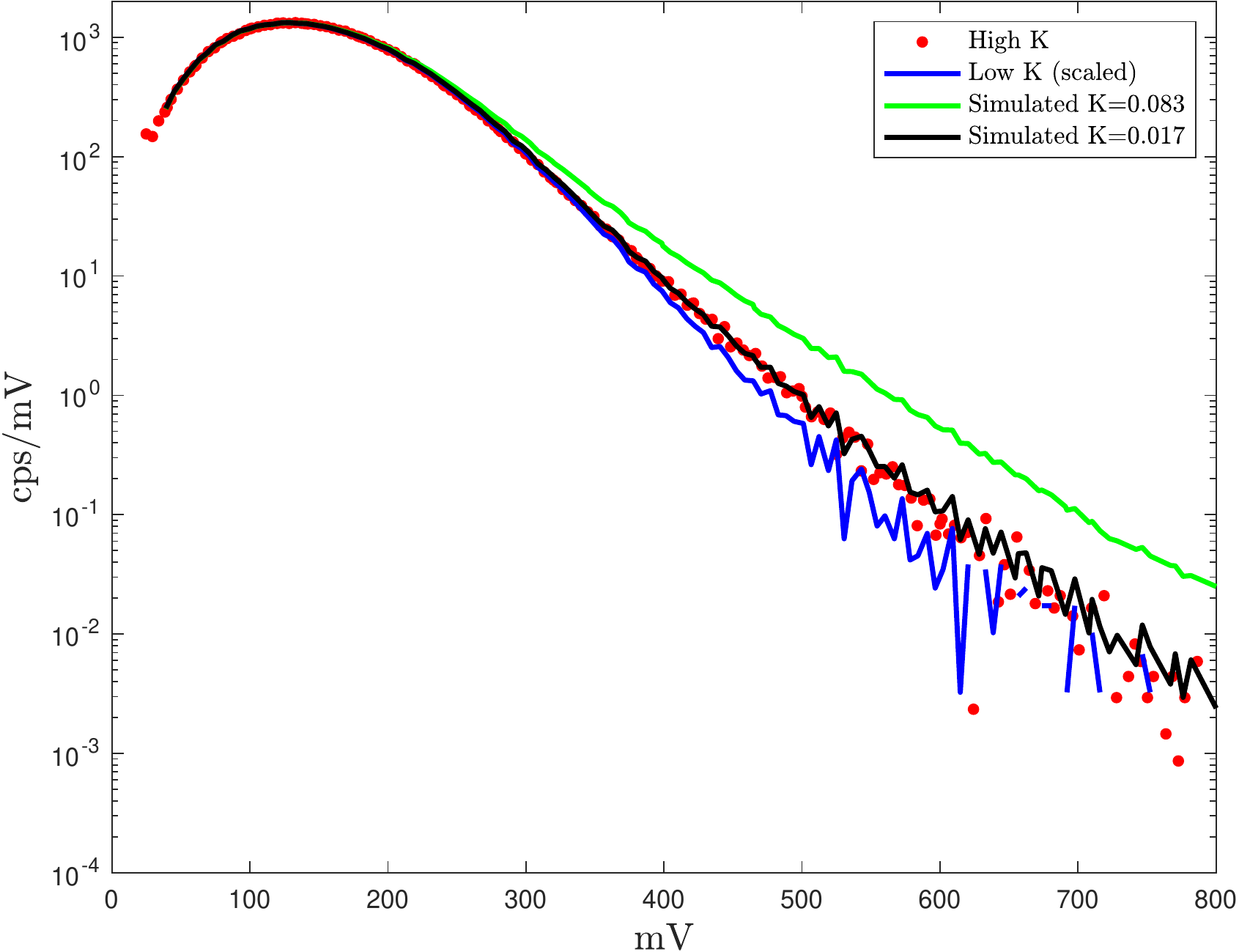}
    \caption{PHD for the low-$K$ measurement scaled to the high-$K$ measurement, compared with simulations for $K=0.083$ and $K=0.017$. The $y$-axis is logarithmic to show clearly the effect of QSA in the larger recorded pulses.}
    \label{fig:datandmodelphds}
\end{figure}

\clearpage
\section{Time Series of Measured Pulses}
\label{timeseriesmethods}
The previous section described a measurement of the energy deposited at the EM by secondary ions. The fundamental quantity of interest was energy/ion, and we used this quantity to constrain the QSA effect. In this section, we will look at a different quantity, the time interval $\Delta t$ between secondary ion arrivals at the EM.  This quantity will allow us to more precisely constrain the statistics of quasi-simultaneous emission and quasi-simultaneous arrival.

\subsection{Model of Secondary Ion Arrivals at the EM}

Ions emitted from the sample are accelerated and travel down the mass spectrometer before they are detected at the EM. During this travel, they may be blocked by the energy slits after the ESA, or at other apertures and slits in the mass spectrometer. To build a useful model, we must link ions measured at the EM (ion arrivals) with ions emitted from the sample (ion emissions). 


Jones et al.\ \cite{jones2020statistical} proposed a two-stage model where the Poisson-distributed quasi-simultaneous emission was followed by a binomial process to create additional multiple ions. The authors provided tabulated values for this ``$K_s^*$'' conditional distribution in a spreadsheet, but it can be calculated analytically (finite sum) as a binomial and Poisson conditional probability distribution:

\begin{align}
    p(x)&=\sum_y p(x|y)p(y)=\sum_{y=0}^{\infty} {y \choose x-y}p^{x-y}(1-p)^{2y-x}\left( \frac{e^{-K}K^y}{y!} \right)\\
    &=(1-p)^{-x}p^x e^{-K}\sum_{y=ceil(x/2)}^x \frac{\left(K(1-p)^2/p \right)^y}{(x-y)!(2y-x)!} \label{eqn:clivepdf}
\end{align}

\begin{figure}
    \centering
    \includegraphics[width=0.3\textheight]{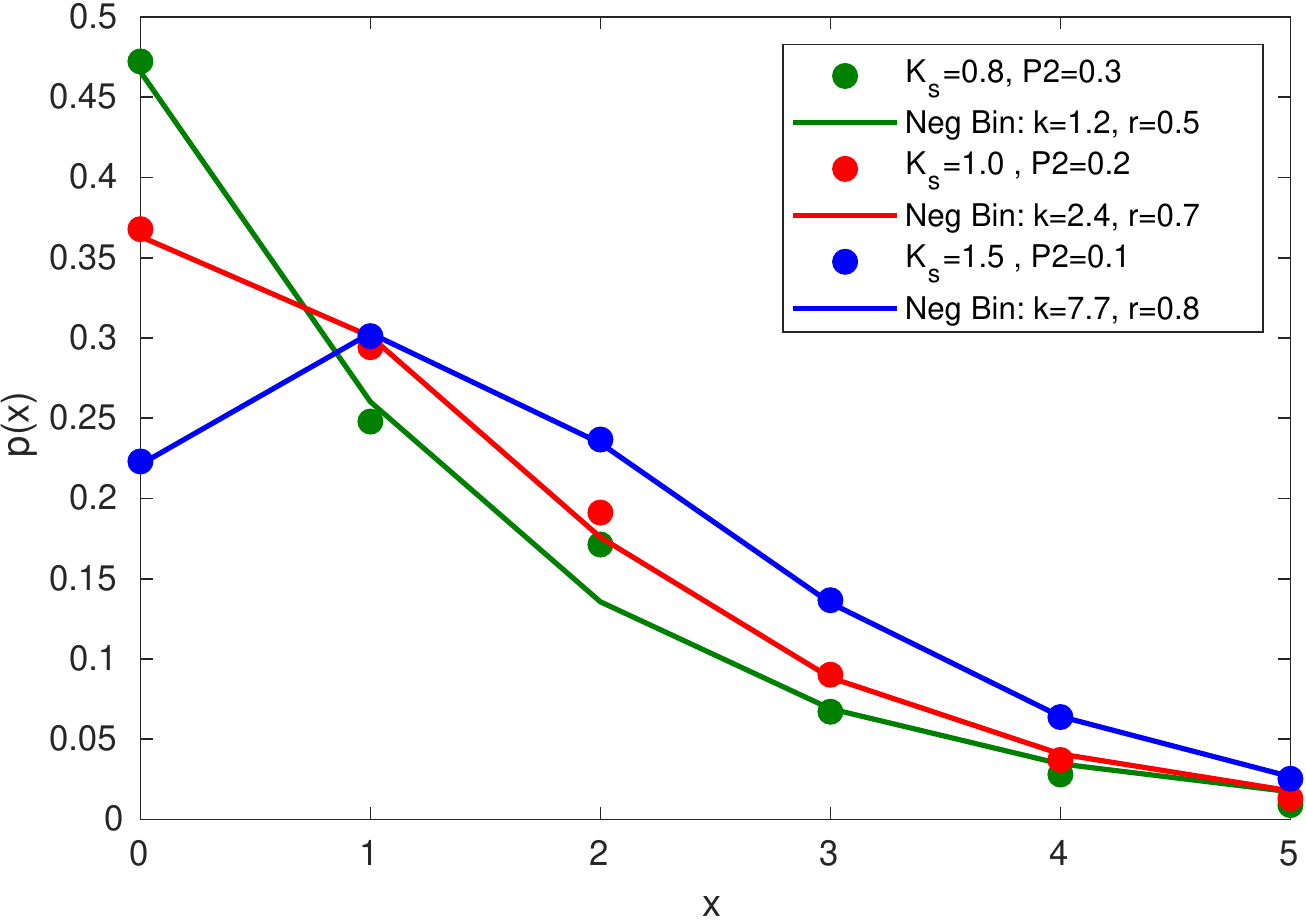}
    \caption{The probability distribution of $x$ multiple arrivals. Solid circles show the probability distribution proposed by \cite{jones2020statistical} for various values of $K_s$ ($K$ in Equation \ref{eqn:clivepdf}) and $P_2$ ($p$ in Equation \ref{eqn:clivepdf}). Lines show the negative binomial approximation to the three different distributions.
    \label{fig:clivepdf}}
\end{figure}

This equation reproduces the values tabulated in the supplementary information of \cite{jones2020statistical}. This probability distribution is well-approximated by a negative binomial distribution as shown in Figure \ref{fig:clivepdf}. However, the distribution in Equation \ref{eqn:clivepdf} suffers from instabilities when $p>K$ and has discontinuities when $x$ is even or odd, so the negative binomial distribution is a more robust statistical model.

In our model, we assume that the initial emission of secondary ions from the sample is described by a negative binomial distribution instead of a Poisson distribution. The negative binomial models the number of failures in a sequence of independent, identically distributed Bernoulli trials (with probability of success $p$) before $r$ successes occur:

\begin{equation}
    P(k) = \binom{k+r-1}{k} p^r(1-p)^k
\end{equation}

The variance of the negative binomial distribution is $\mu(1+\mu/r)$. In the limit where $r \rightarrow \infty$, the negative binomial distribution reduces to the Poisson distribution with mean $\mu$. The negative binomial allows for a mean that is different from its variance, which allows for a more accurate model of the QSA effect in the case that it is over- or under-dispersed compared to the Poisson distribution. In this paper, we will use the negative binomial distribution, which is consistent with the distribution proposed by \cite{jones2020statistical}.

If $r$ can take on non-integer positive values, we replace the binomial coefficient by the gamma function:

\begin{equation}
    P(k) = \frac{\Gamma(k+r)}{\Gamma(k+1)\,\Gamma(r)} p^r (1-p)^k 
\end{equation}

The probability $p(y)$ that $y$ ions arrive at the detector (after emission from the sample surface and passage through the apertures) is given by the marginal probability distribution:

\begin{equation}
    p(x)=\sum_y p(x|y)p(y)
    \label{eqn:conditional_prob}
\end{equation}

That is, the distribution of $y$ is equal to the product of the probability distribution of $x$ given $y$, multiplied by the probability distribution of $y$, summed over all possible values of $y$. For example, the probability that zero ions make it to the detector ($x=0$) is equal to: the probability that zero ions are emitted by the initial negative binomial process ($y=0$) multiplied by the probability that zero of these ions make it through the apertures, plus the probability that one ion is emitted ($y=1$) multiplied by the probability that it does not make it through the apertures, plus the probability that two ions are emitted ($y=2$) multiplied by the probability that neither of these makes it through the apertures, etc.

The distribution of $p(y)$ is the negative binomial distribution parameterized by $p$ and $r$. The conditional distribution $p(x|y)$ is the binomial distribution (with probability $q$) where the number of independent experiments is equal to $y$, the number of ions emitted from the sample by the negative binomial process. Plugging these two distributions into Equation \ref{eqn:conditional_prob}:

\begin{align}
    NBB(x) &\equiv p(x)=\sum_y p(x|y)p(y)\\
    &=\sum_{y=0}^{\infty} \frac{\Gamma(y+r)}{\Gamma(y+1)\,\Gamma(r)} p^r (1-p)^y  {y \choose x} q^x(1-q)^{y-x}\\
    &=\frac{p^r q^x}{\Gamma(r)} \sum_{y=x}^{\infty} \frac{\Gamma(y+r)(1-p)^y}{\Gamma(y+1)} {y \choose x} (1-q)^{y-x}
    \label{eqn:conditional_binomial_poisson}
\end{align}

We calculate $NBB(x)$ by summing terms until the last term changes the sum by no more than some threshhold value (typically $10^{-7}$). Since the distribution proposed by \cite{jones2020statistical} (Equation \ref{eqn:clivepdf}) can be approximated by the negative binomial distribution (Figure \ref{fig:clivepdf}), it can also be approximated by $NBB(x)$. Therefore, the model we propose here is consistent with the mechanism proposed by \cite{jones2020statistical}.


In our model we assume that $q$ is the probability that the ion passes through the ESA and other apertures. Since the ion's energy affects the arrival times we will model, we express $q$ as the product of two probabilities: $q=q_e q_a$, where $q_e$ is the probability that the ion has the right energy to pass through the ESA, and $q_a$ is the probability that the secondary ion will pass through all of the other slits, apertures, and energy slit to arrive at the detector. This probability $q_e$ can be calculated from the energy scan ($F(E)$ vs.\ $E$) shown in Figure \ref{fig:energyscan}.

\begin{equation}
    q_e=\frac{\int_{E_{min}}^{E_{max}}F(E)dE}{\int_{-\infty}^{+\infty}F(E)dE}
\end{equation}

As we have no way to calculate $q_a$ \textit{a priori}, we will vary $q_a$ so that our model fits the measured time differences in the arrival pulses.


\subsubsection{Energy Distribution of Secondary Ions}

We model the probability distribution of the kinetic energy of secondary ions $P(\Delta E)$ following \cite{wittmaack2014comprehensive}: 

\begin{equation}
\label{eqn:energydist}
    P(\Delta E) = \frac{2 (\Delta E)^2}{E_s^2(1+\Delta E/E_s)^3}\left(1+erf\left(\frac{E+0.5\omega}{\sqrt{2}\sigma}\right)\right) \left(1-erf\left(\frac{E+0.5\omega}{\sqrt{2}\sigma}\right)\right)
\end{equation}

where $E_s$ is the surface binding energy, and $\omega$ and $\sigma$ are characteristic widths of the rectangular box and Gaussian which are convolved to make the shape of the energy spectrum.

\begin{figure}
    \centering
     \includegraphics[width=0.5\textheight]{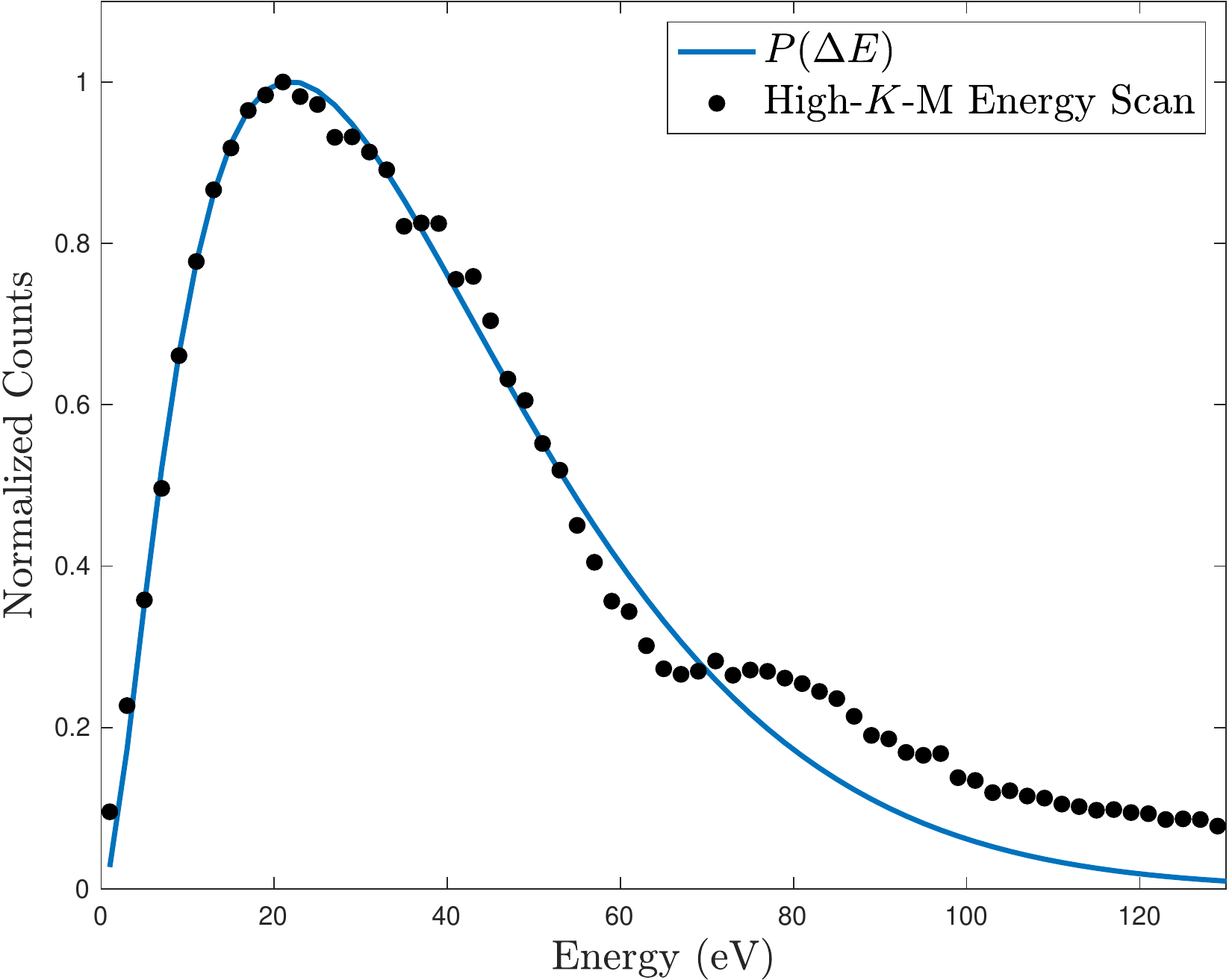}
    \caption{Energy distribution and model fit (Equation \ref{eqn:energydist}) of secondary ions of $^{16}$O$^-$ in magnetite (High-$K$-M dataset) measured on the UH Cameca ims 1280.}
    \label{fig:energyscan}
\end{figure}

The measured energy spectrum for the ``High-$K$-M'' magnetite standard measured in our analysis is shown in Figure \ref{fig:energyscan}. We measured the energy spectrum by scanning a narrow energy slit and recording the counts of $^{16}$O$^-$ at the EM.  We fit this energy spectrum using $E_s=15.7$eV, $\omega=29.1$ eV, and $\sigma=45.4$~eV. Most of the secondary ions are fit well by this model, but the tail of the distribution above 60~eV is a relatively poor fit. The higher energy secondary ions may be affected by physical processes not accounted for in Equation \ref{eqn:energydist}, or the way this spectrum was measured (scanning the energy slit after the ESA) may be less accurate at the higher energies. This inaccuracy will not be a significant source of error in modeling the QSA phenomenon, as the important features of the secondary ion energy distribution are modeled accurately.

\subsubsection{Detailed Model Description}

We model the emission and detection of ions in a secondary ion mass spectrometer as follows:
\begin{enumerate}
    \item Primary ions impact the sample surface. 
    \item A number of secondary ions are emitted from the sample, described by a Poisson or Negative Binomial distribution.
    \item The kinetic energy $\Delta E$ of the emitted ions before they are accelerated are randomly drawn from the energy scan (Figure \ref{fig:energyscan}).

    
    \item The secondary ions are accelerated by the extraction voltage $E_{ext}$ and travel the distance of the secondary flight path of the mass spectrometer $D$, with total kinetic energy given by $E_{ext}+\Delta E$. The travel time of the (nonrelativistic) ions is given by:
    \begin{equation}
        \Delta t=\frac{D}{\sqrt{2(E_{ext}+\Delta E)/m}}
    \end{equation}
    
    \item The secondary ions pass through the electrostatic analyzer which acts as a bandpass filter, where only energies from $E_{min}$ to $E_{max}$ are permitted.
    \item The secondary ions have some probability $q_a$ of being stopped by the apertures.
    \item The secondary ions arrive at the detector.
\end{enumerate}

An analytical solution for the arrival times at the detector is not possible given the form of the probability density functions and energy dependence of the secondary ions, so we use numerical techniques to simulate $5 \times 10^6$ ions for each measurement condition. We have included our Matlab code to perform these simulations in the supplementary information. 

\subsubsection{Comparison to Previous Work}
The effect of quasi-simultaneous ions for Steps 1, 2, and 7 above, has been calculated previously (e.g., \cite{slodzian2001precise}). Equivalently, if one sets the length of the secondary flight tube $D$ equal to zero, an analytical solution is easy to calculate---it is the formalism described previously \cite{slodzian2004qsa}.

\begin{figure}
    \centering
    \includegraphics[width=0.5\textwidth]{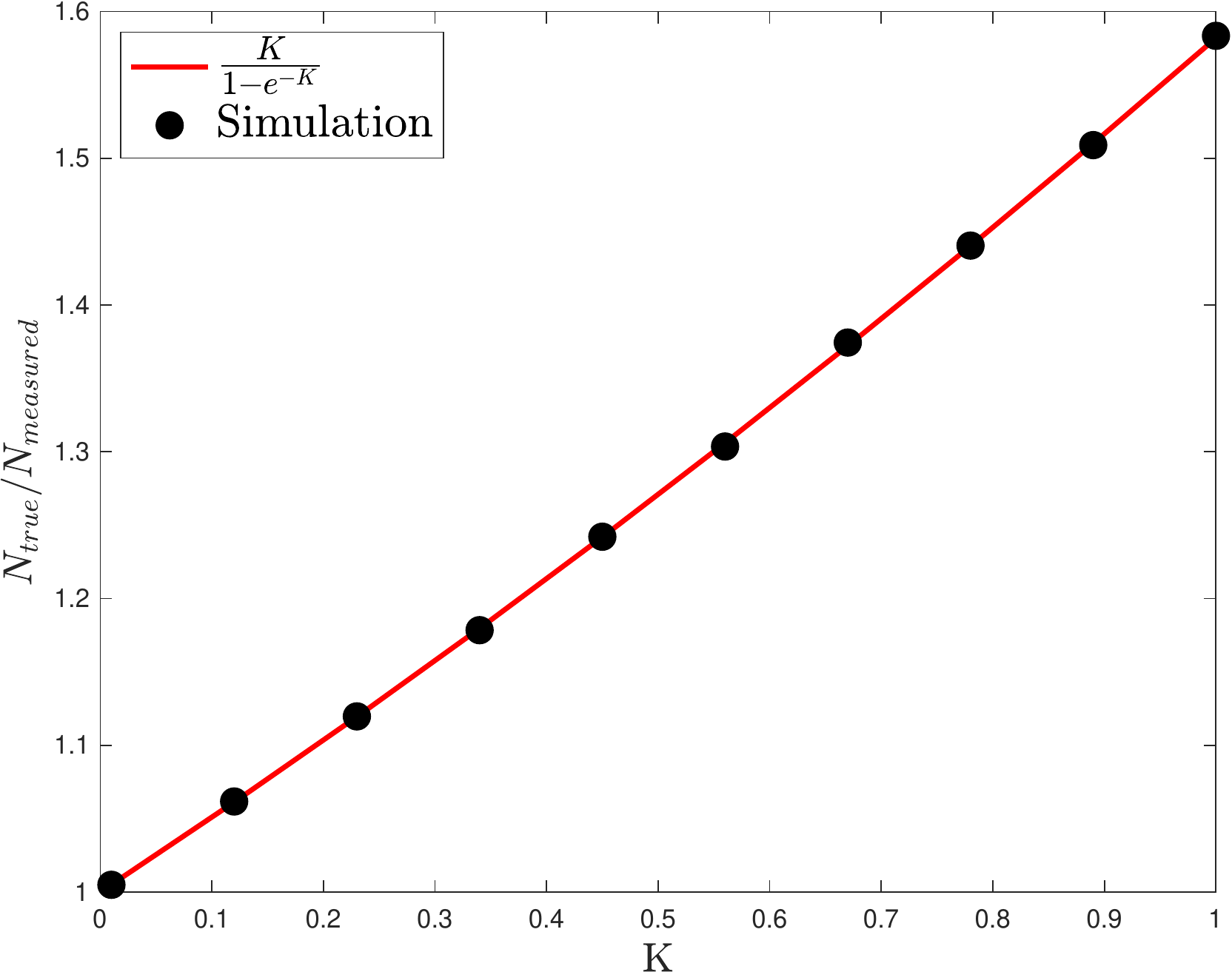}
    \caption{Numerical simulation of the QSA correction $N_{true}/N_{measured}$ vs. $K$, the secondary to primary count rate ratio, compared to the true value for zero secondary ion flight distance.}
    \label{fig:slodz}
\end{figure}

Using the numerical procedure outlined above, we set the length of the secondary ion flight tube equal to zero to reproduce this treatment of the QSA effect. As shown in Figure \ref{fig:slodz}, the numerical procedure reproduces $N_{true}/N_{measured}$ for various values of $K$.

Arrival times between pulses without the QSA effect will be distributed according to the exponential probability density function with rate parameter $\lambda$. We estimate $\lambda$ from the data (assuming the QSA pulses will not affect $\lambda$ significantly). This expected distribution is shown by the blue dashed curves in Figure \ref{fig:pulseseparationsnomodel}. 

\subsection{Measurements of Secondary Ion Arrivals at the EM}
\label{measurementsofsecondaryionarrivals}
We directly measured arrival times of secondary ions ($^{16}$O$^-$ from chromite and magnetite, except $^{17}$O$^-$ was measured in the ``Low-$K$'' dataset) on the University of Hawaii Cameca ims 1280 ion probe. Some important measurement conditions were as follows---primary current: 1.7~nA, primary high voltage: 10~kV, exit slit width: 172~$\mu$m, mass resolving power: 7076, entrance slit: 69~$\mu$m, field aperture: 5000~$\mu$m.

We recorded output voltage from the preamplifier as a seamless data log using a digital oscilloscope (Tektronix TDS5104B). Voltage of the output was sampled every 0.4 nanoseconds for 500,000 points. We recorded $\sim$4000 pulses for each of the four different measurement conditions given in Table \ref{tab:meas_conditions}. 

\begin{table}
\label{tab:meas_conditions}
\caption{Parameters for time series measurements of arrival pulses.}
\begin{center}
\begin{tabular}{l c c c c}
Name & $K$ & Phase & CPS & Energy Slit\\\hline 
Low-$K$ & $<$0.001 & Chromite & 4.9$\times$10$^5$&75~eV\\
Mid-$K$ & 0.152 & Chromite & 9.5$\times$10$^5$&75~eV\\
High-$K$ & $>$0.152 & Chromite & 1.05$\times$10$^6$&Open\\
High-$K$-M & $>$0.152 & Magnetite & 9.5$\times$10$^5$&Open\\
\end{tabular}
\end{center}
\end{table}

\begin{figure}
    \centering
    \includegraphics[width=0.5\textwidth]{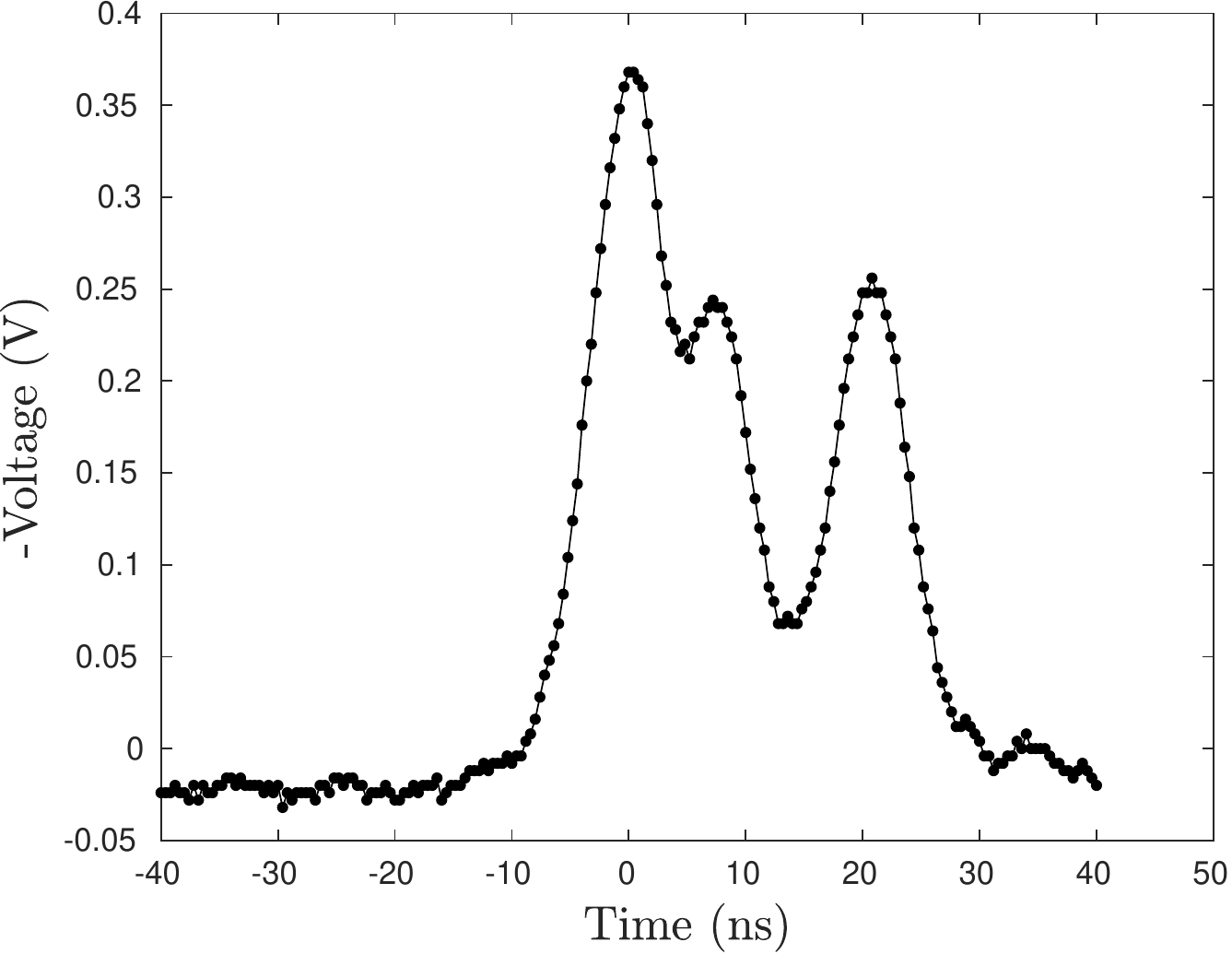}
    \caption{Three close pulses as measured by a digital oscilloscope on the first dynode of the EM (from the High-$K$ measurement).}
    \label{fig:closepeaks}
\end{figure}

In these voltage time-series datasets we identified peaks using Matlab's \texttt{findpeaks} then fit them with Gaussians, which was a good approximation of the peak shape. Peak widths are $\sim$5.0$\pm$1.5 nanoseconds and are determined by the electron cascade dynamics (likely the dominant source), the response of the preamp, and the response of the oscilloscope. We are able to reliably find peaks with centers as close together as $\sim$5~nanoseconds. Three close peaks are shown in Figure \ref{fig:closepeaks}.

\begin{figure}
    \centering
        \includegraphics[height=0.25\textheight]{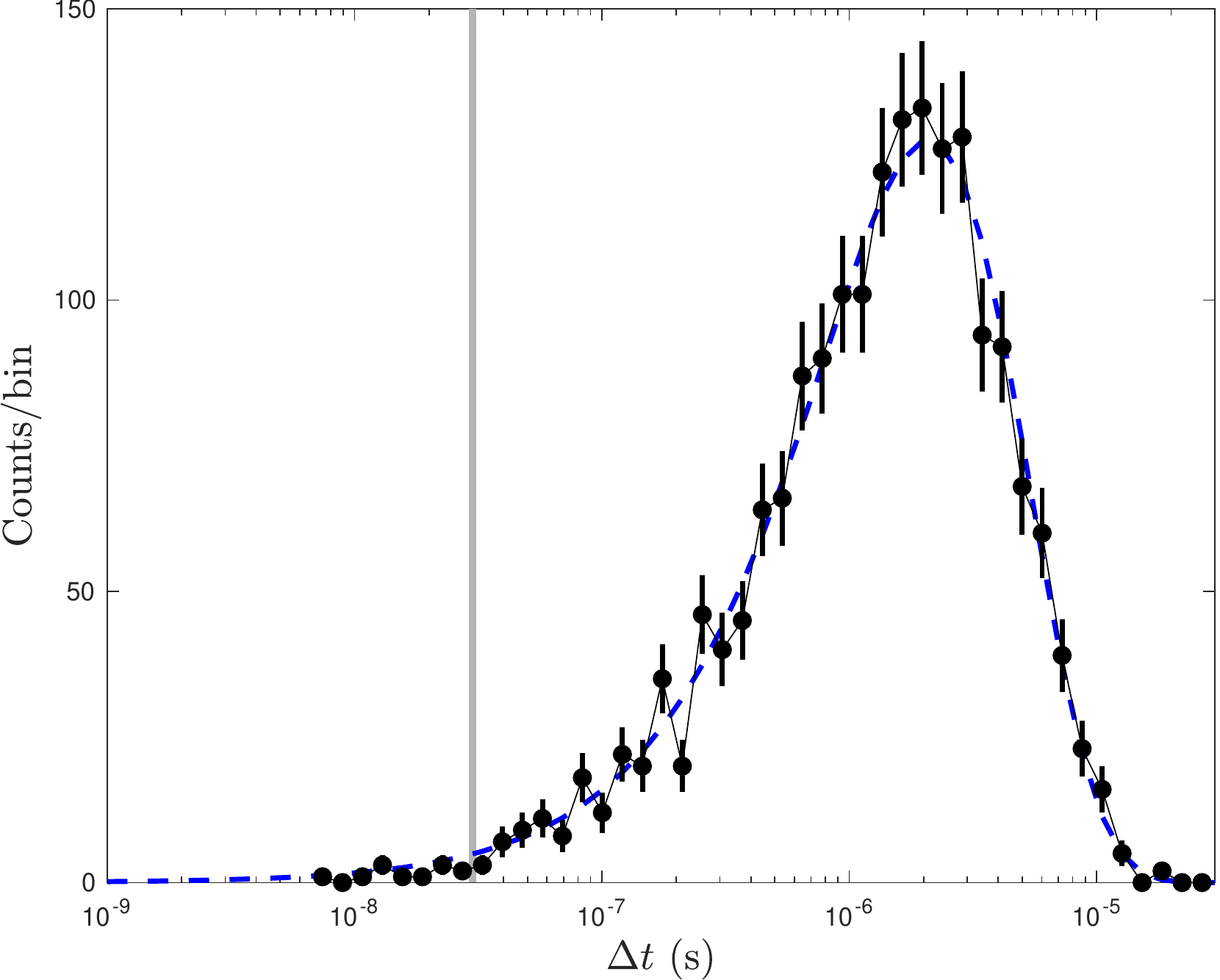}
    \includegraphics[height=0.25\textheight]{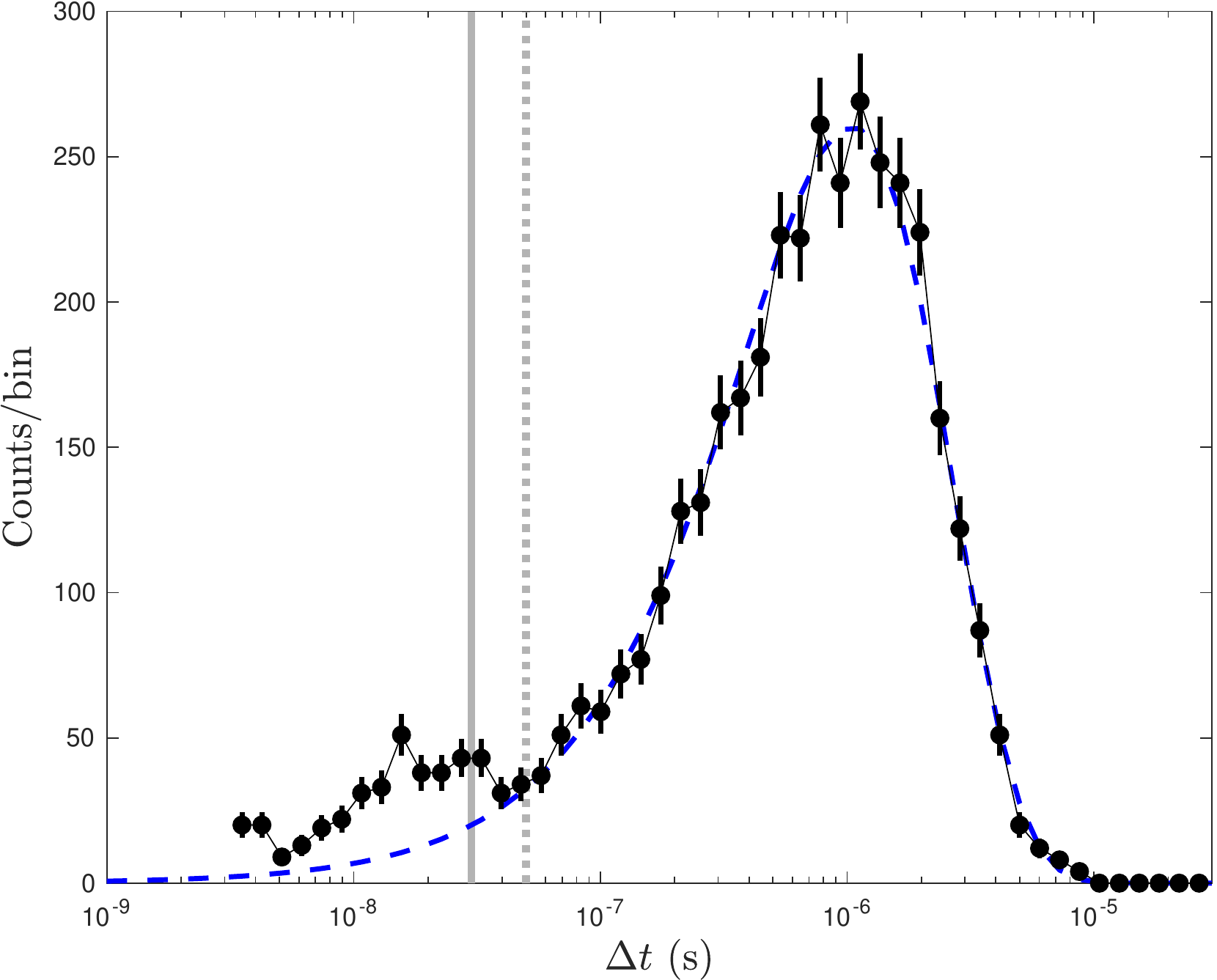}
    \includegraphics[height=0.25\textheight]{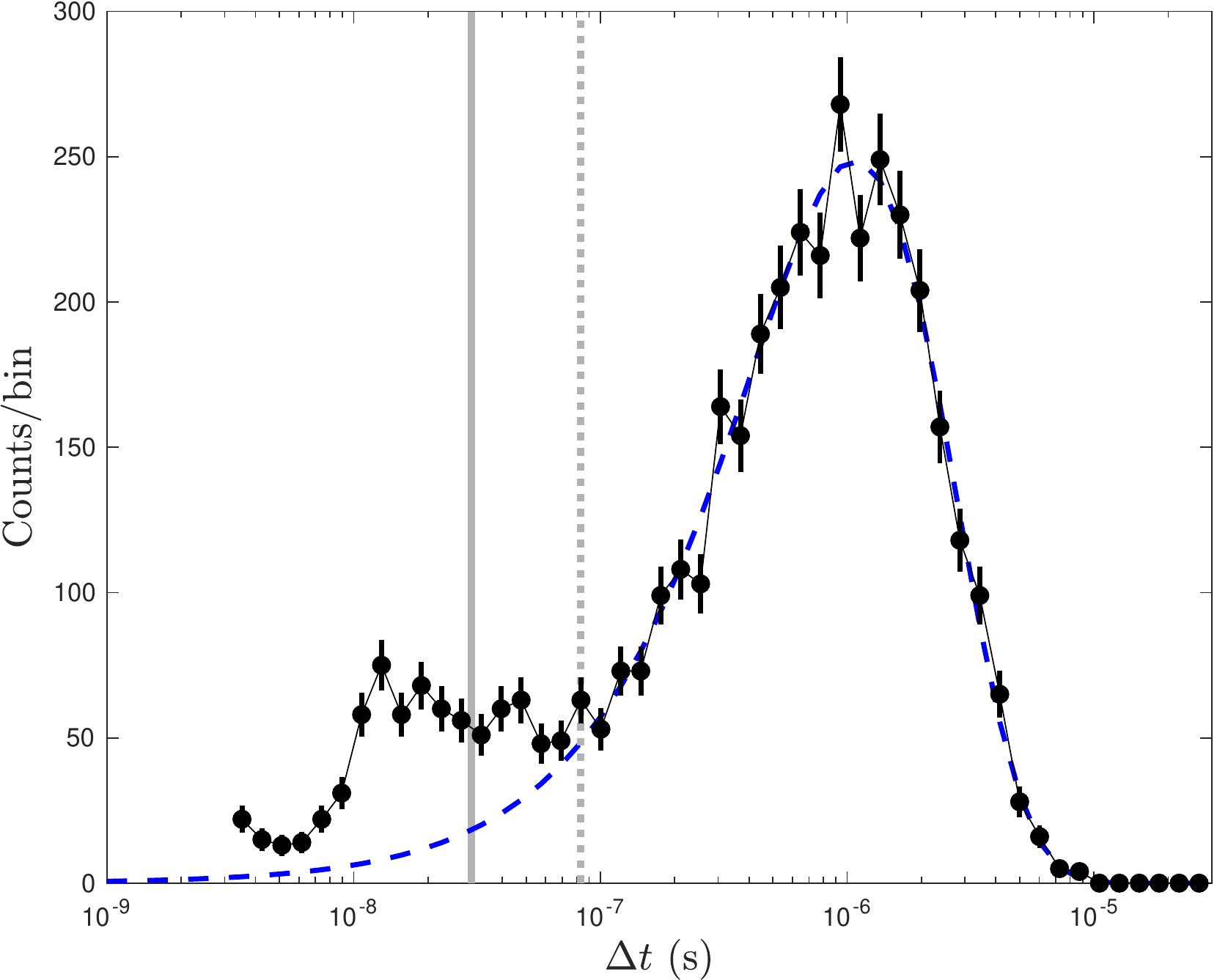}
        \includegraphics[height=0.25\textheight]{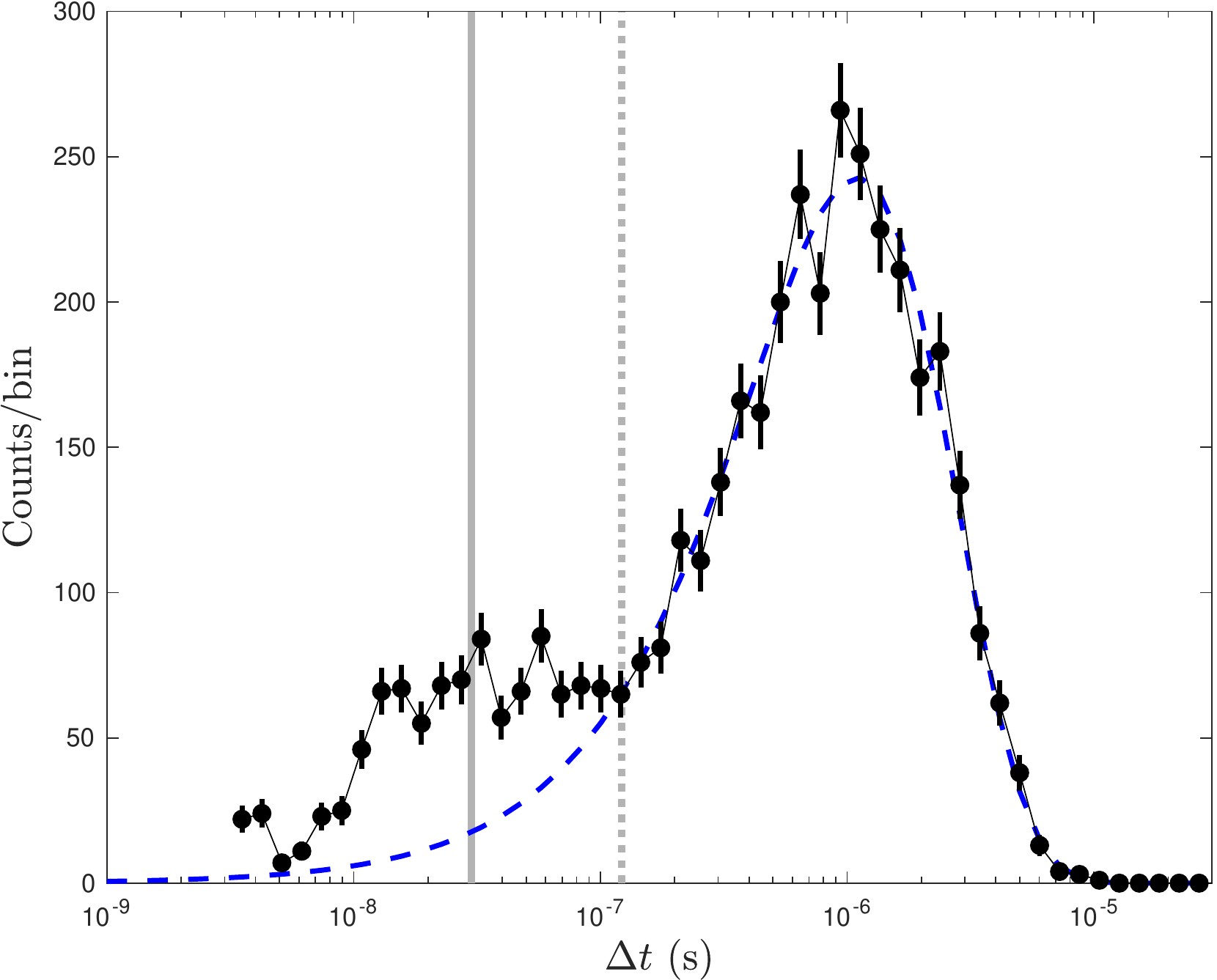}

    \vspace{-1pc}
    \caption{Histograms of the time between pulses at the first dynode of the EM. The measured pulse separation histograms are shown by filled black circles. Top-left: low-$K$, top-right: Mid-$K$, lower-left: High-$K$, lower-right: High-$K$-M. Error bars are $\pm$ the square root of the number of counts in each bin. The blue dashed curve is the exponential distribution with rate parameter $\lambda$ equal to the mean count rate for that measurement.  Vertical solid grey line indicates the EM deadtime (30~ns). Vertical dashed grey line indicates the largest $\Delta t$ when quasi-simultaneous arrivals are significant ($\Delta t_{\text{thresh}}$). Pulses arriving at smaller $\Delta t$ than the EM deadtime (left of the solid grey line) will be counted as a single pulse.} \label{fig:pulseseparationsnomodel}
\end{figure}

We calculated the numerical derivative of these peak times which yields the time between arrivals of pulses $\Delta t$. Then we calculated histograms of the $\Delta t$ values with logarthmic-spaced bins to highlight the QSA counts at small $\Delta t$. Histograms for the four conditions described in Table \ref{tab:meas_conditions} are shown in Figure \ref{fig:pulseseparationsnomodel}, compared to the expected distribution of arrival time differences in a Poisson process (exponential distribution of $\Delta t$) without QSA ions.

\begin{figure}
    \centering

        \includegraphics[height=0.2\textheight]{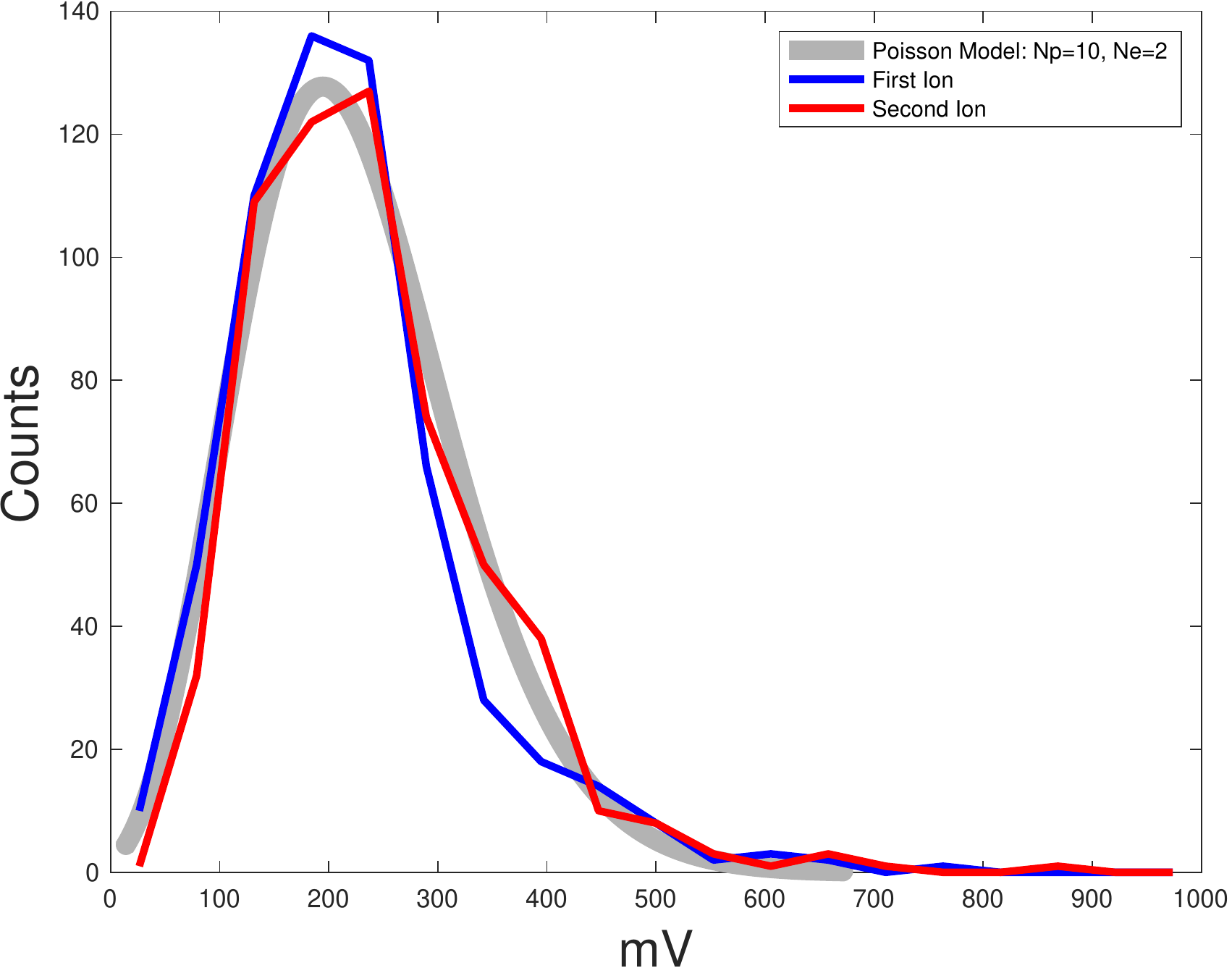}
    \includegraphics[height=0.2\textheight]{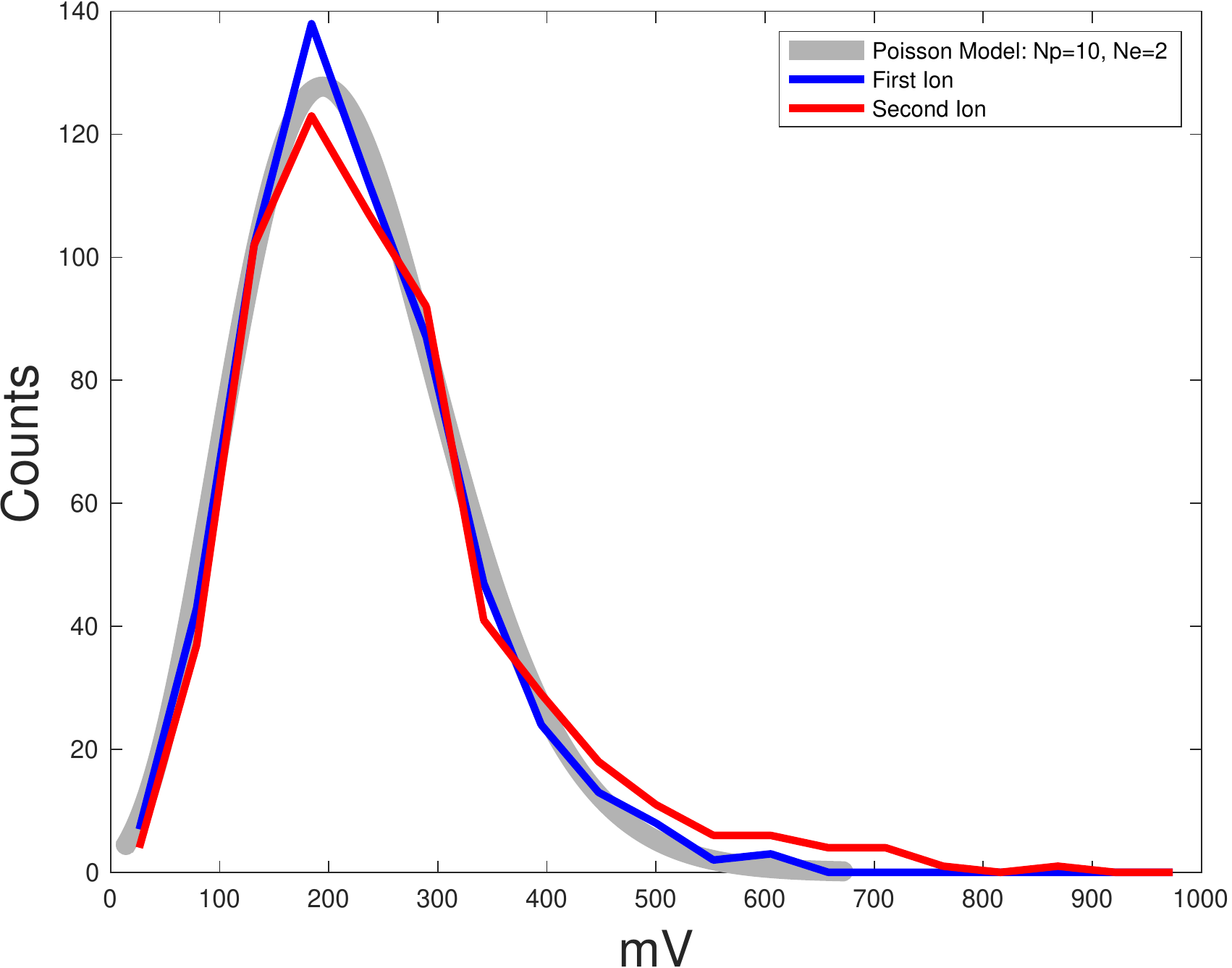}

    \vspace{-1pc}
    \caption{Pulse-height distributions for first and second ions in double arrivals for the High-$K$ (left) and High-$K$-M (right) measurements. \label{fig:QSA_PHDs}}
\end{figure}

The pulse-height distribution of QSA ions can tell us if the EM has time to fully recover before the next pulse in a QSA pair arrives. We calculated the peak heights of QSA pairs and calculated the PHDs of the first-arriving and second-arriving ions (Figure \ref{fig:QSA_PHDs}). There is no significant difference between the PHDs of the first and second arriving ions, so we conclude that the EM fully recovers even between closely arriving ions.

\subsection{Results: Statistical Distribution of QSA Ions}
We calculate the number of multiple arrivals in the Mid-$K$, High-$K$, and High-$K$-M data sets to see how well these pulses follow a Poisson process. For the time series we identify consecutive pulses that are more closely placed than $\Delta t_\text{thresh}$. Quasi-simultaneous arrival does not necessarily mean quasi-simultaneous emission (secondary ions that were emitted simultaneously from a single incident primary ion). Some pulses will arrive closer than $\Delta t_\text{thresh}$ because of the exponentially distributed time between emissions given by the Poisson process. This is the blue dashed-line curve in Figure \ref{fig:pulseseparationsnomodel}, and the correction for these missed counts is just the normal deadtime correction calculation. We correct for the expected double arrivals from normal Poisson emission using the cumulative exponential distribution function: $1-e^{-\lambda \Delta t_\text{thresh}} \approx$10\% of the total pulses, and assume that if one ion is emitted with low energy ($\Delta E = 0$~eV), and the next is emitted with high energy ($\Delta E = 100$~eV), it might be possible for the fast ion to catch up with the slow ion and arrive at the detector closer than $\Delta t_\text{thresh}$. We calculate that the difference in time between two such ions is $\sim$100~ns which is much smaller than the characteristic time between secondary ion emissions ($\sim$1~$\mu$s), so we conclude that velocity dispersion of non-quasi-simultaneous emitted ions is not a major source of quasi-simultaneous arrivals on the EM. 

The number of excess double, triple, etc.\ arrivals was calculated for the Mid-$K$, High-$K$, and High-$K$-M data sets, and the expected number from the exponential distribution (blue curve in Figure \ref{fig:pulseseparationsnomodel}) was subtracted. We compare these measured multiple arrivals with what is expected from a Poisson distribution with a mean value $K_0$. We calculated the best fit $K_0$ for each dataset as: $K_0=2P(2)/P(1)$, where $P(2)$ is the measured fraction of double arrivals and $P(1)$ is the measured fraction of single arrivals.

\begin{figure}
    \centering
    \includegraphics[width=0.4\textwidth]{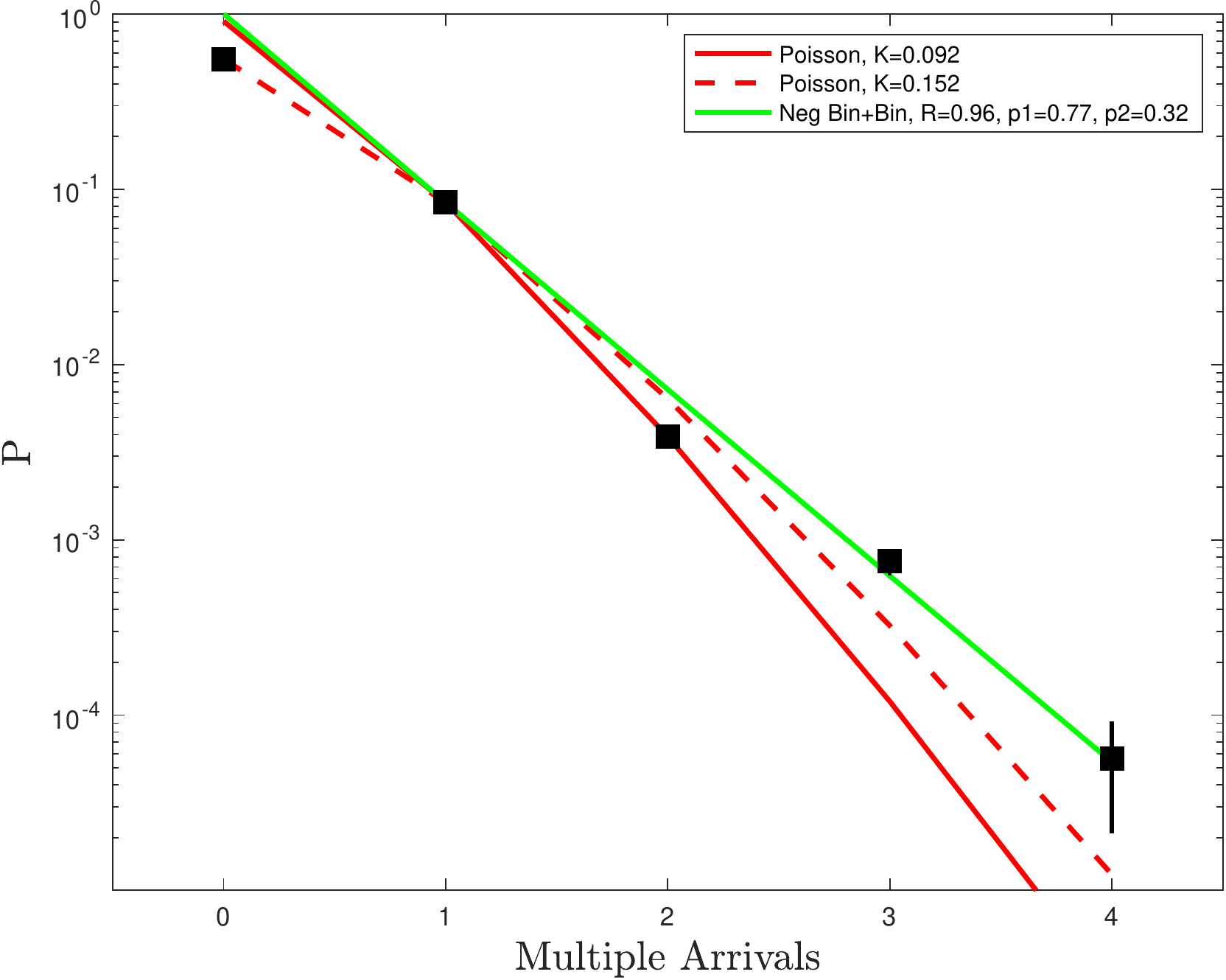}
    \includegraphics[width=0.4\textwidth]{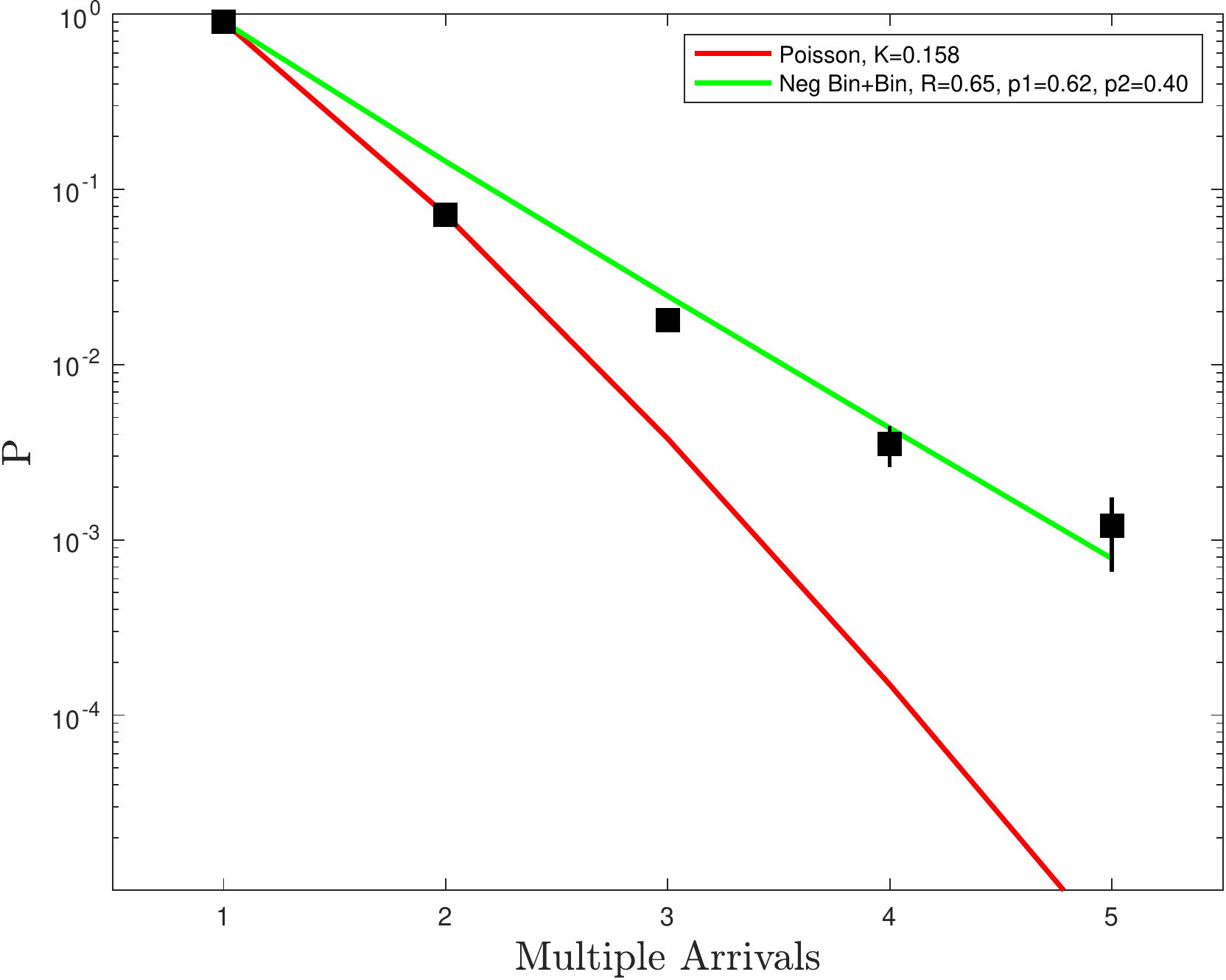}
    \includegraphics[width=0.4\textwidth]{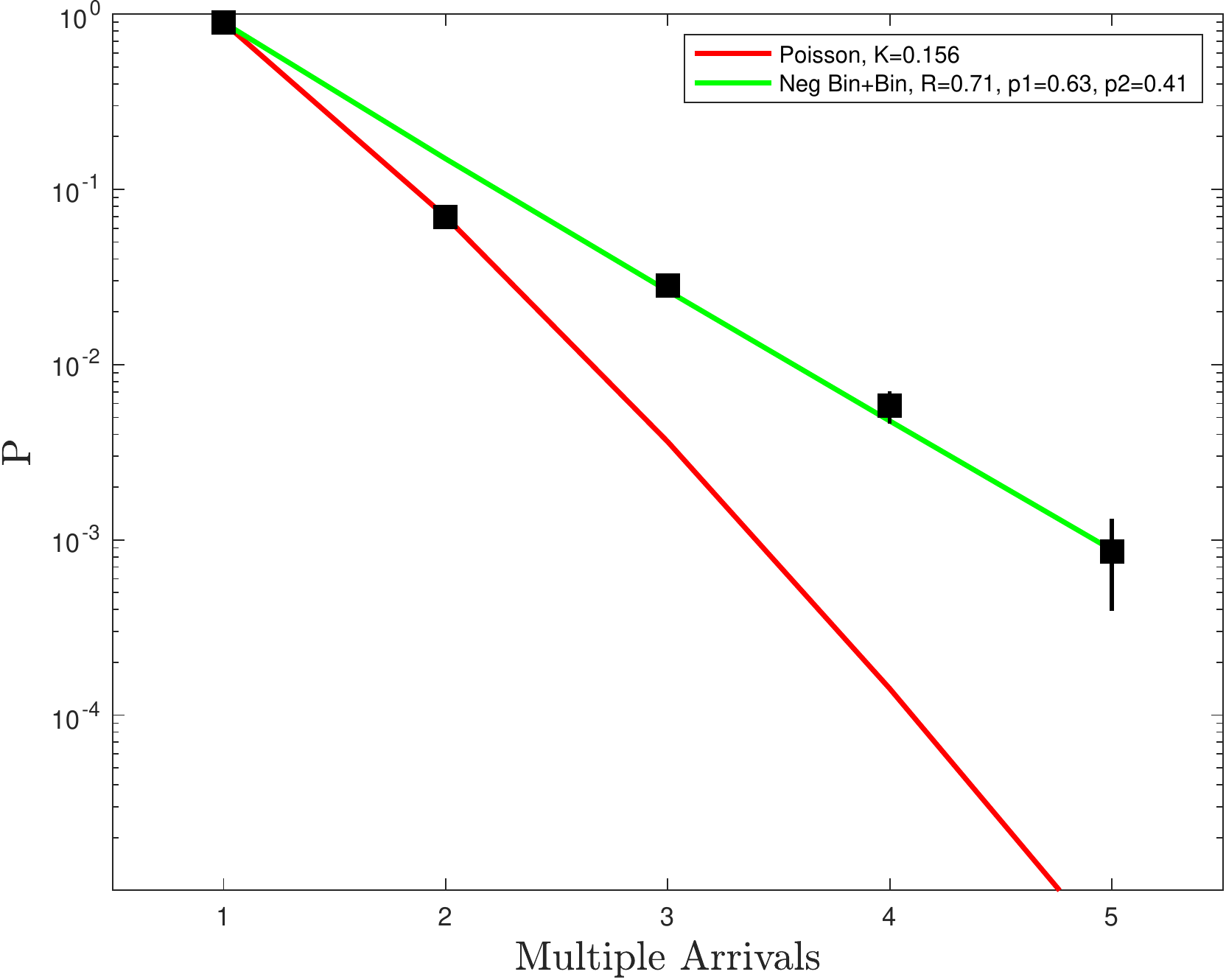}
\caption{Probability distribution of excess multiple arrivals for the Mid-$K$ (top-left), High-$K$ (top-right), and High-$K$-M (bottom) measurements. Solid red line is a Poisson fit using only the single and double arrivals. Red dashed line in Mid-$K$ measurement is the Poisson distribution with mean equal to the measured secondary-to-primary ratio (0.152). Green solid line is the $NBB$ probability distribution fit.  \label{fig:multiplearrivals}}
\end{figure}

The primary to secondary ion count rate ratio was determined to
be 0.152 for the Mid-$K$ measurement. The primary beam
currents for the high-$K$ and high-$K$-M
measurements were below detection limit of the primary Faraday
cup, so only a lower bound on this ratio was determined for these two datasets.. The low-$K$ measurement had too few double counts to be useful for evaluation of the statistical distribution of QSA ions. For the Mid-$K$ dataset, we can test the accuracy of the Slodzian et al. \cite{slodzian2001precise} model, where multiple arrivals are Poisson distributed with mean equal to the secondary to primary ratio. The Poisson probability distribution, with $\lambda=0.152$, is compared to the actual distribution of secondary ions in Figure \ref{fig:multiplearrivals}.

A Poisson model that fits single and double counts for QSA \cite{slodzian2004qsa} undercounts triple, quandruple, etc. arrivals, as shown by the difference between the red curve and the measured data points in Figure \ref{fig:multiplearrivals} . The difference between the Poisson model and triple+ arrivals is worse for higher $K$. A fit with a $NBB$ distribution underestimates double arrivals but is more accurate for triple+ arrivals. The true distribution of multiple arrivals does not appear to follow an analytical probability distribution.



\subsection{Effect of Changing Energy Window and Secondary Ion Energy Distribution}
\label{effectofchangingenergy}

\begin{figure}[!htpb]
    \centering
    \includegraphics[width=0.4\textwidth]{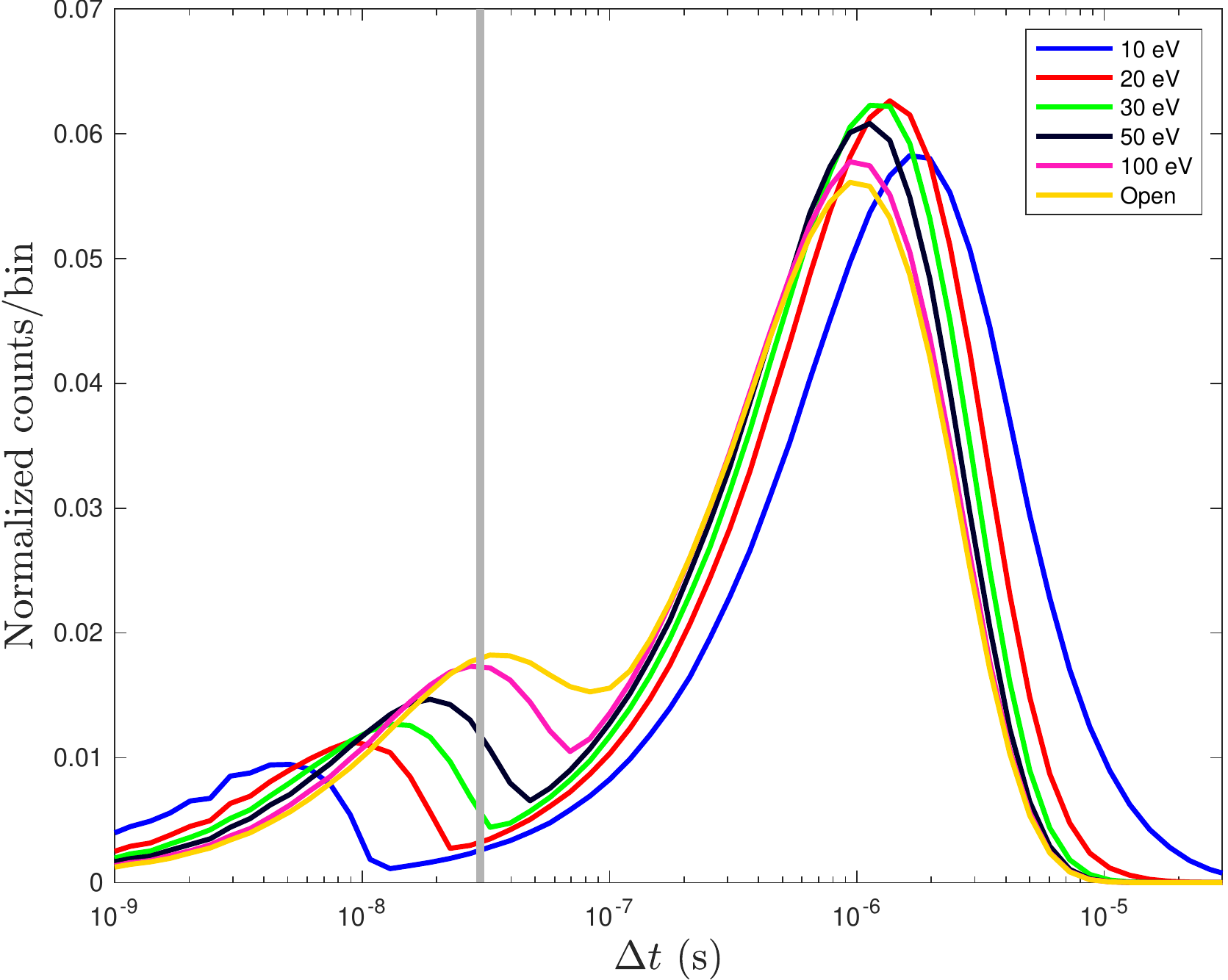}
    \includegraphics[width=0.4\textwidth]{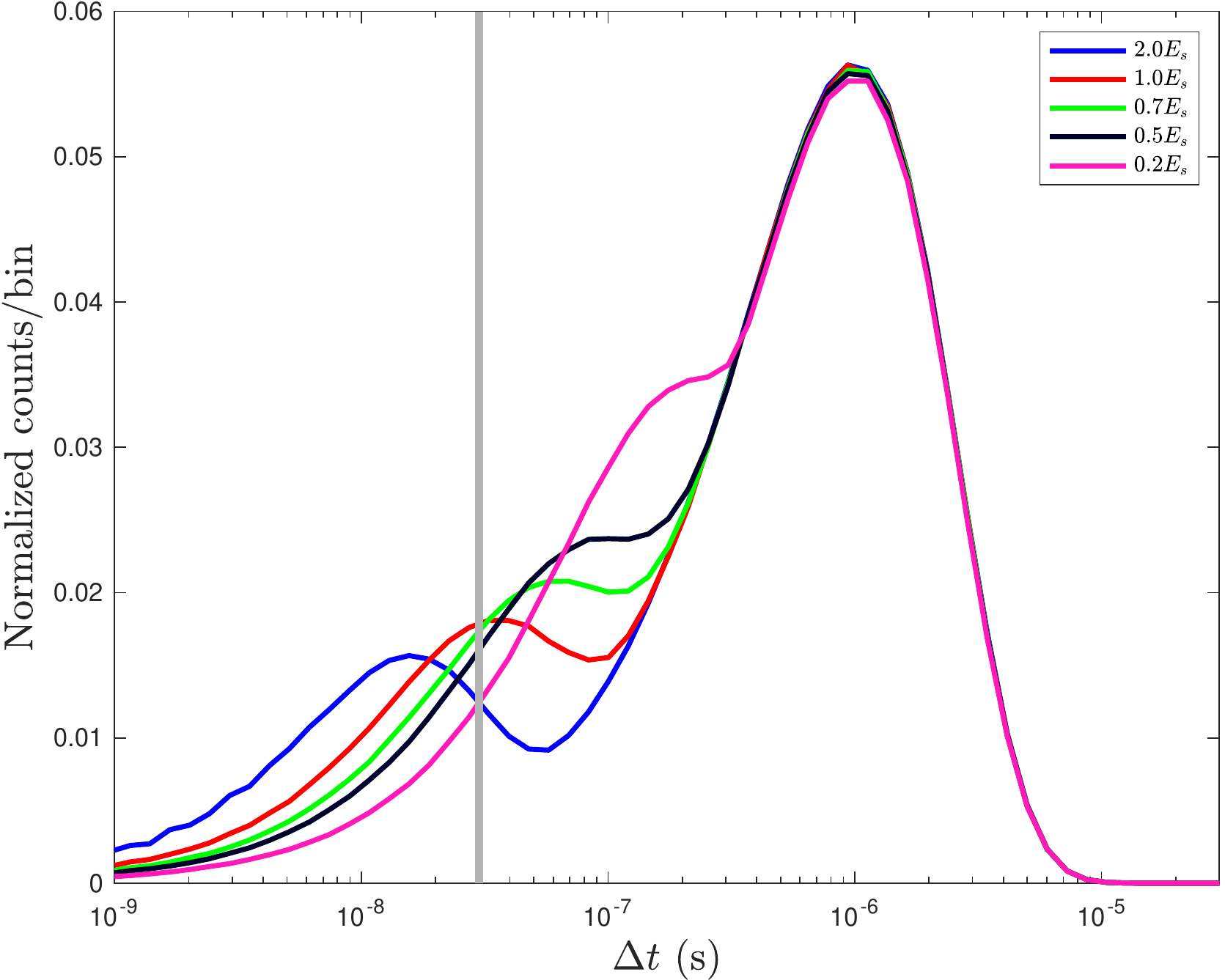}
    \caption{Simulations of time between pulses showing the effect of changing the energy window (left) and changing $E_s$ in the energy distribution for secondary ions (right). Multiple arrivals follow the NBB fit to the High-$K$-M data and the secondary ion energy distribution is initially assumed to be equal to that measured for the High-$K$-M dataset. \label{fig:energywindowvary}}
\end{figure}

The position and width of the energy slit, a bandpass filter, will affect the energy distribution of secondary ions. This, in turn, will affect the secondary-ion flight time, and how close in time the secondary ions reach the EM. A change in the binding energy will also affect the secondary ion kinetic energy, flight times, and distribution of arrival times of secondary ions. We modeled the dependence of the pulse separation distributions on energy slit and secondary ion surface binding energy (results are shown in Figure \ref{fig:energywindowvary}). For these simulations we used the $NBB$ fit to multiple arrivals in the High-$K$-M measurement (Figure \ref{fig:multiplearrivals}).


For our measurements, we recorded the energy slit setting for each of the four measurements shown in Table \ref{tab:meas_conditions}. We only measured the energy distribution of secondary ions for the Mid-$K$ setting, but this distribution may change as the measurement conditions change. To account for these changes, we adjusted the surface binding energy $E_s$ so that the simulation best matches the data. The probability that the secondary ion makes it through the apertures, $q_a$, was set equal to one as a smaller number did not significantly improve the model fit to the data.

\subsection{Results: Simulation of QSA Effect}
\begin{figure}
    \centering
    \includegraphics[width=0.4\textwidth]{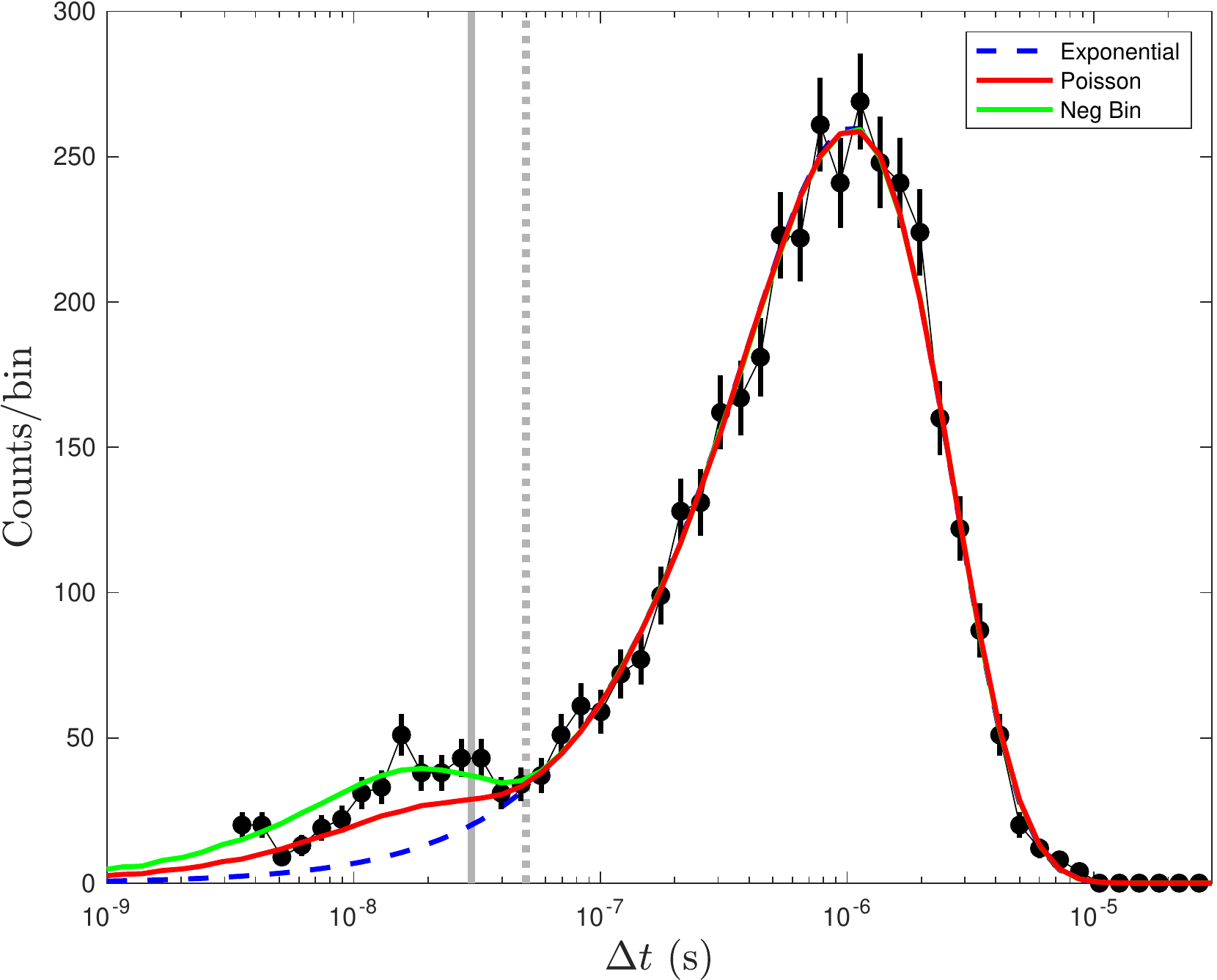}
    \includegraphics[width=0.4\textwidth]{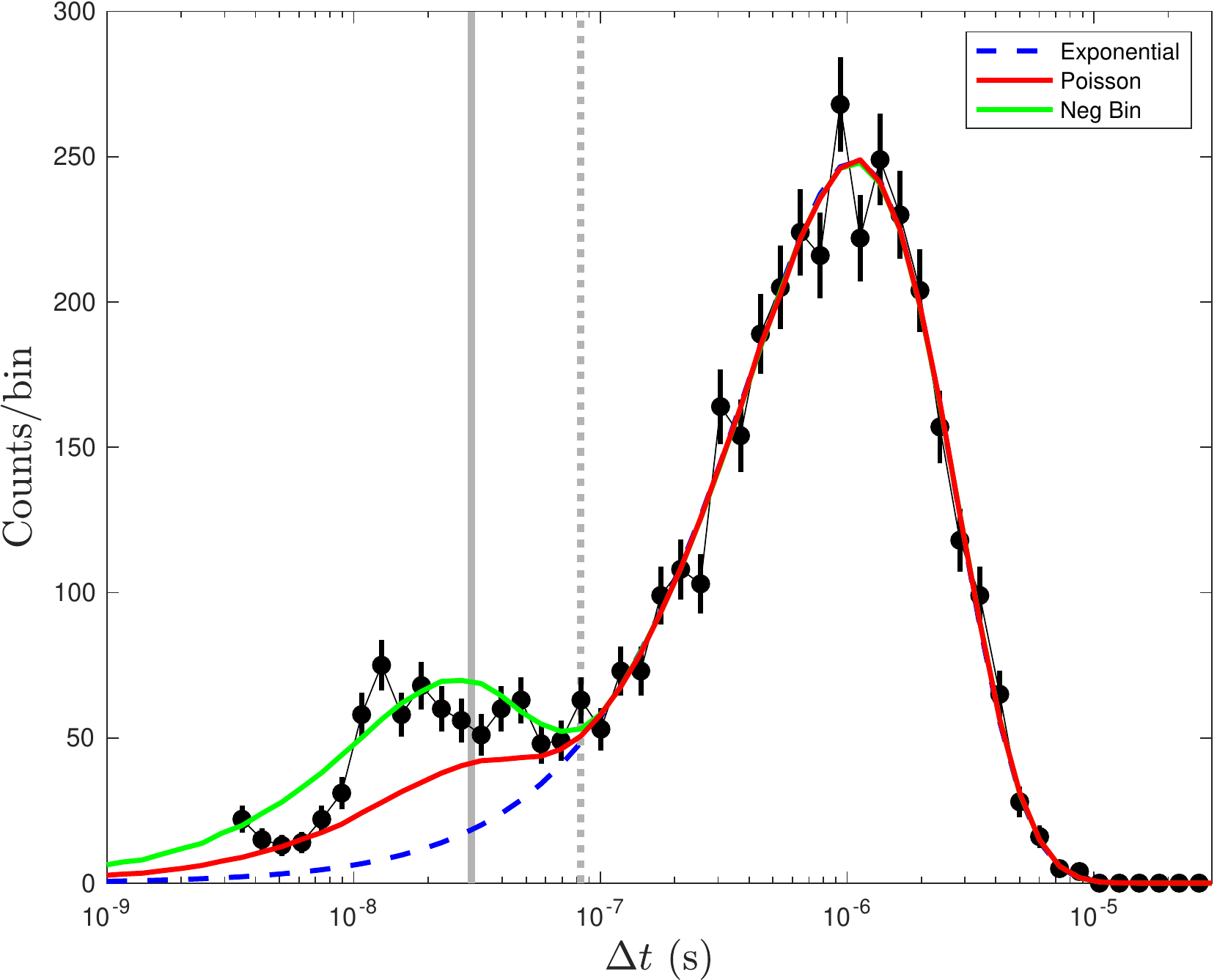}
    \includegraphics[width=0.4\textwidth]{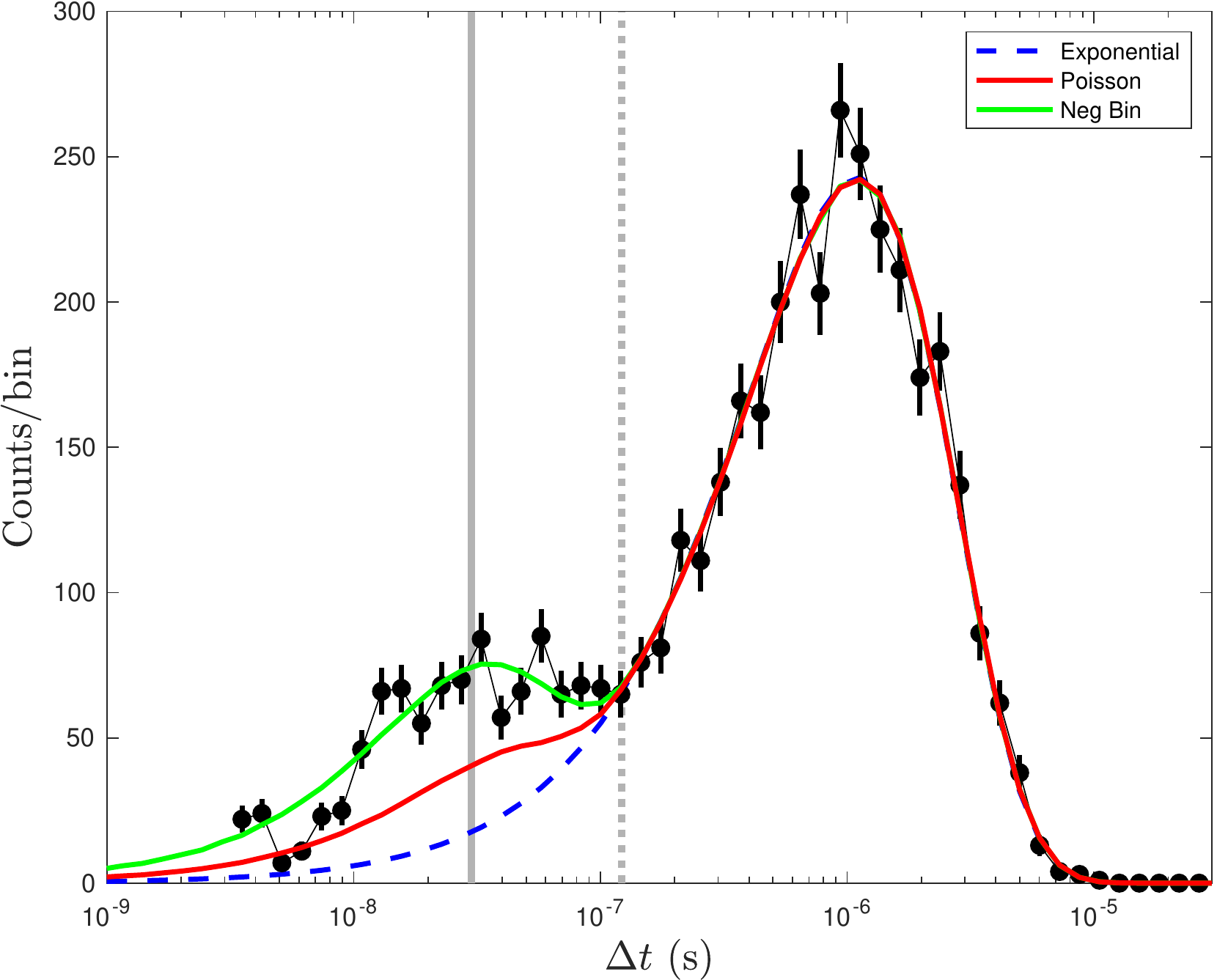}
    \caption{Histograms of the time between pulses at the first dynode of the EM. Blue dashed lines are the same as Figure \ref{fig:pulseseparationsnomodel} and represent the no-QSA case. a) Mid-$K$ ($K$=0.152) with the energy distribution measured at the High-$K$-M setting but with $E_s$ multiplied by 2.1, energy window = 75~eV, and multiple-arrival threshhold = 50~ns. b) High-$K$ with the energy distribution measured at the Huge-$K$-M setting but with $E_s$ multiplied by 1.3, energy window open, and multiple-arrival threshhold = 83~ns. c) High-$K$-M with energy window open, and multiple-arrival threshhold = 122~ns.} \label{fig:pulseseparations}
\end{figure}

The results of the simulation for the Mid-$K$, High-$K$, and High-$K$-M data sets are shown in Figure \ref{fig:pulseseparations}. The surface binding energy $E_s$ (Equation \ref{eqn:energydist}) was adjusted for the Mid-$K$ and High-$K$ data to better fit the QSA peak. We assume that counts that arrive closer than the EM deadtime are counted as a single pulse, so that the final ion counts need to be multiplied by a correction factor greater than one. These correction factors are given in Table \ref{tab:correctionfactors}.
  
The important observations of Figure \ref{fig:pulseseparations} are:
\begin{enumerate}
    \item Quasi-simultaneously emitted ions may be emitted with different energies (Equation \ref{eqn:energydist}) and drift apart from each other during travel through the instrument, so that their arrival time separation is larger than the deadtime. That is, quasi-simultaneous emission does not always result in quasi-simultaneous arrival. For smaller geometery ion probes like the Cameca 7f, secondary ions may not have time to drift apart enough to be detected separately.
    \item The $NBB$ probability distribution is a better fit to the data but overestimates pulses closer than ten nanoseconds. However, this may be a bias in the data analysis---it is more difficult to detect close pulses in our data (Figure \ref{fig:closepeaks}) and some may be counted as one pulse.
    \item Chromite (High-$K$ and Mid-$K$) required a higher surface binding energy $E_s$ than magnetite (High-$K$-M), by factors of 1.3 and 2.1 for High-$K$ and Mid-$K$, respectively.  This is likely because these two materials have very different surface sputtering properties under the primary Cs$^+$ beam, and that the electron flood gun was used for chromite but was not used for magnetite. Changes in the secondary ion energy distribution may cause the quasi-simultaneously emitted ions to drift apart outside the deadtime window, and change the correction factor.
    \item Opening the energy slit allows more secondary ions through, increasing the fraction of counts lost to QSA, but some of these ions of different energies may drift apart outside the deadtime window, decreasing the fraction of counts lost to QSA.
    \item In the Poisson model of \cite{slodzian2004qsa}, the correction factor estimated by the measured $K$ (0.152) is too large ($K=0.123$ yields the appropriate correction, Table \ref{tab:correctionfactors}). 
\end{enumerate}


\begin{table}
\caption{Correction factors ($N_{true}/N_{measured}$) for the simulations and data shown in Figure \ref{fig:pulseseparations}. \textit{True} is the correction from the data (black filled circles in Figure \ref{fig:pulseseparations}), \textit{Neg Bin Model} is the green curve, \textit{Poisson Model} is the red curve, \textit{Poisson Expr} is Equation \ref{eqn:poissonexactcorr} with $K$ given in Figure \ref{fig:multiplearrivals}, $K_{fit}$ is the $K$ value in Equation \ref{eqn:poissonexactcorr} that gives the \textit{True} correction factor.  \label{tab:correctionfactors}}
\begin{center}
\begin{tabular}{l c c c c c}
Name & True & Neg Bin Model & Poisson Model & Poisson Expr & K$_{fit}$\\\hline
Mid-$K$ & \textbf{1.093} & 1.119 & 1.074 & 1.079 & 0.119\\
High-$K$ & \textbf{1.138} & 1.180 & 1.091 & 1.116 & 0.199\\
High-$K$-M & \textbf{1.135} & 1.164 & 1.080 & 1.115 & 0.193\\
\end{tabular}
\end{center}
\end{table}

\clearpage

\section{Discussion}
Our measurements show that the QSA effect depends in a complicated fashion on the secondary ion energy distribution, the secondary ion flight length, the secondary ion acceleration voltage, and the electron multiplier deadtime. The NanoSIMS PHDs (Figure \ref{fig:datandmodelphds}) and the Mid-$K$ pulse-separation measurements show that the oversimplified Poisson model of \cite{slodzian2004qsa} overcorrects the QSA effect. However, other SIMS measurements summarized in \cite[][and references therein]{jones2020statistical} show that often the Poisson model undercorrects the QSA effect. The statistical distribution of multiple arrivals and emissions (Figure \ref{fig:multiplearrivals}) may vary with secondary ion species ($^{16}$O$^-$ from chromite and magnetite was measured here for all but the low-$K$ measurement, where $^{17}$O$^-$ was measured). Under the conditions measured here, the distribution of secondary arrivals is signficantly overdispersed compared to Poisson. An overdispersed distribution results in more lost ions to QSA \cite{jones2020statistical}. The energy distribution of secondary ions and ion probe analytical conditions may result in a different fraction of ions drifting apart enough to be detected individually. A shorter EM deadtime, longer secondary flight path, and broader secondary ion energy spectrum results in fewer counts lost to QSA. These combined effects may result in either an overprediction of the Poisson QSA correction for some measurements and an underprediction in others. This makes it impossible to accurately correct for the QSA effect analytically. Even a semi-empirical approach, where the QSA correction is determined over a range of analytical conditions, will be inaccurate because these combined effects do not smoothly vary. 



The best protocol is to measure appropriate standards, acquired with similar analytical protocols and with similar secondary ion count rates, and normalize the unknown measurements to these standards.

In geochemistry and cosmochemistry, isotope ratios are usually expressed with the most abundant isotope in the denomiator. If both isotopes are measured with EMs, the major isotope is more likely to be affected by QSA, since the QSA effect increases with the secondary/primary ion ratio. For example, in the ratio $^{18}$O/$^{16}$O, the QSA effect on $^{16}$O is $\sim$500 times larger than on $^{18}$O. We will assume that the QSA effect can be ignored on the numerator isotope.  

Cosmo/geochemists are often interested in deviations of isotope ratios from known terrestrial values. The fundamental quantity of interest is $R_u$ --- the isotope ratio of the unknown sample to the same ratio in some standard. For example, oxygen isotopes are often expressed as parts-per-thousand (per mil) deviations from a standard ocean water value (Vienna standard mean ocean water or VSMOW). We will call the normalizing ratio
$R$. For $^{18}$O/$^{16}$O, $R$=0.0020052 for VSMOW. However, since measured SIMS ratios will differ from the true ratio due to instrumental fraction effects \cite{eiler1997sims}, we cannot just calculate $R_u=r_u/R$. We measure our unknown isotope ratio ($r_u$) and the same isotope ratio measured in a standard ($r_s$) of similar composition as our unknown, but with known isotope ratio ($R_s$). Various processes in the instrument causes the measured value of the standard $r_s$ to be different from its true value $R_s$. We assume that the relative error in the unknown is the same as the sample, because the standard and unknown have similar compositions. The true ratio of the unknown to $R_u$ is then \cite{kita2010high}:

\begin{equation}
    R_u=\dfrac{r_u}{R\left(\dfrac{r_s}{R_s}\right)}=\dfrac{r_u R_s}{r_s R}
\end{equation}

This ratio, expressed in $\delta$ units (per mil), is:

\begin{equation}
\label{eqn:delta}
   \delta = \left( \dfrac{r_u R_s}{r_s R} -1 \right)1000
\end{equation}

In the cosmochemistry literature, an alternative expression for the $\delta$ values is often used \cite{rollion2011determination,villeneuve2019high}:

\begin{equation}
   \delta' = \delta_{unknown}+\left(\delta_{standard,measured}-\delta_{standard,true}\right)
\end{equation}

This expression for $\delta'$ is described as an approximation to Equation \ref{eqn:delta} (e.g. a ``first order'' approximation \cite{villeneuve2019high} for $\delta$ without stating what quantity is used for such a series expansion, and why it is justified to drop higher order terms). In the following we will calculate exactly what is assumed in this approximation, using the definitions of $r_u$, $r_s$, $R_t$, and $R$ defined previously:

\begin{align}
   \delta' &= \delta_{unknown}+\left(\delta_{standard,measured}-\delta_{standard,true}\right)\\
   &=\left(\dfrac{r_u}{R}-1\right)1000 + \left(\dfrac{r_s}{R}-1\right)1000 - \left(\dfrac{R_s}{R}-1\right)1000\nonumber\\
   &= \left( \dfrac{r_u+r_s-R_s}{R} -1 \right)1000\nonumber\\
   &= \left( \dfrac{r_ur_s+r_s^2-R_sr_s}{r_s R} -1 \right)1000\nonumber\\
   &= \left( \dfrac{r_u\left(r_s+\dfrac{r_s^2}{r_u}-\dfrac{R_sr_s}{r_u}\right)}{r_s R} -1 \right)1000\nonumber\\
   &= \left( \dfrac{r_uR_s}{r_s R}\left(\dfrac{r_s}{R_s}+\dfrac{r_s^2}{r_uR_s}-\dfrac{r_s}{r_u}\right) -1 \right)1000
\end{align}

For $\delta$' to be equal to $\delta$, we must have:

\begin{align}
    \dfrac{r_s}{R_s}+\dfrac{r_s^2}{r_uR_s}-\dfrac{r_s}{r_u}&=1\\
    \dfrac{r_s}{R_s}\left(1+\dfrac{r_s}{r_u}-\dfrac{r_s}{r_u}\dfrac{R_s}{r_s} \right)&=1
\end{align}

If the measured standard ratio $r_s$ is equal to the true standard ratio $R_s$ then this equation holds true. However, if it is not true, then $\delta$' will differ from $\delta$.

\begin{figure}
    \centering
    \includegraphics[width=0.7\textwidth]{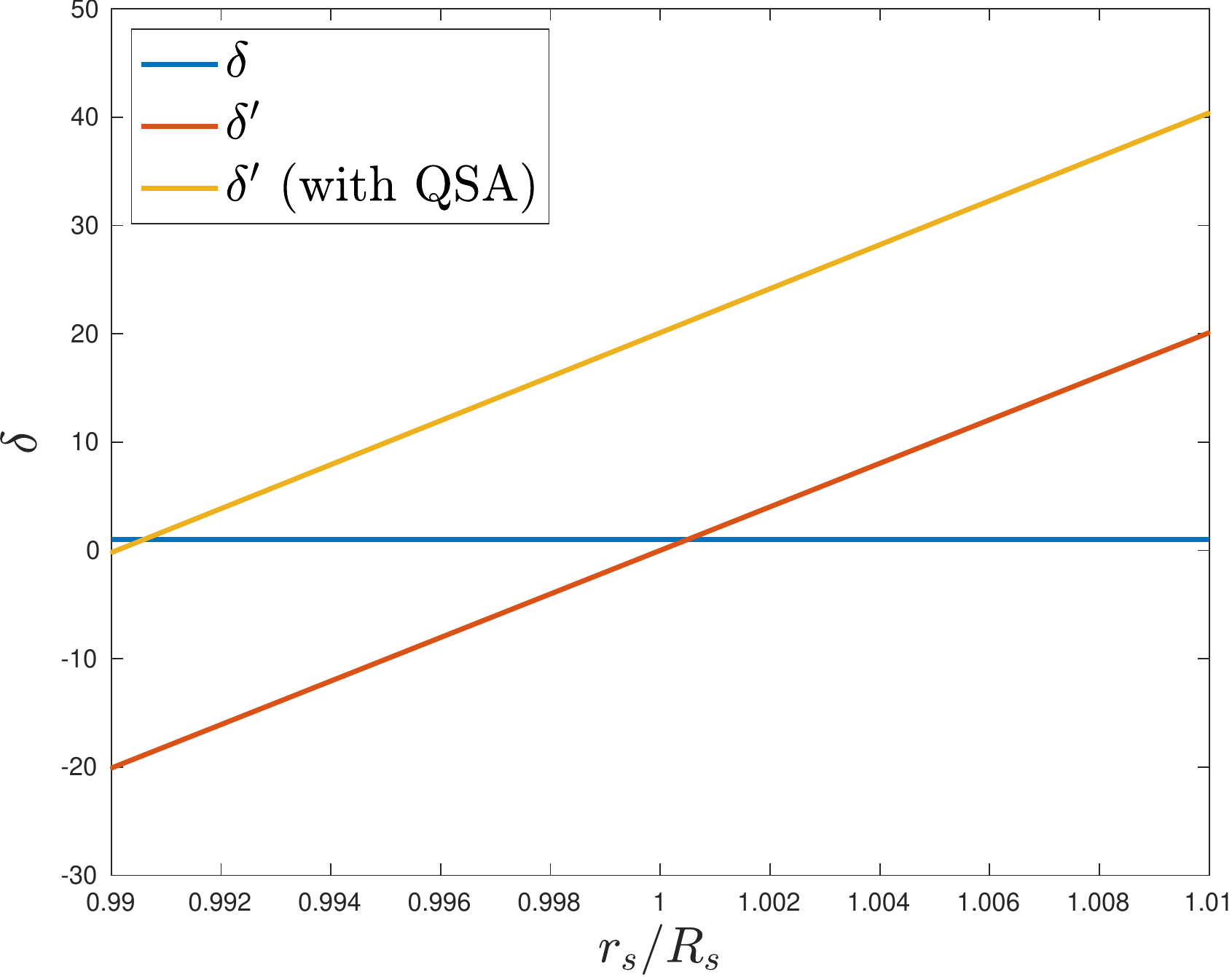}
    \caption{Values of $\delta$, $\delta$', and $\delta$' with a QSA factor of 1.01 for a range of values of the measured-to-true isotope ratio in a standard.} \label{fig:qsadeltafac}
\end{figure}

The other benefit of using the expression for $\delta$ instead of $\delta$' is that the measured ratios $r_u$ and $r_s$ are only used in the ratio $r_u/r_s$. This means that systematic errors that are equal multiplicative factors on both $r_u$ and $r_s$ divide out, whereas they do not in the calculation of $\delta$'. If the standard and unknown are measured under the same conditions (count rates, energy slits, etc.) and have similar secondary ion energy distributions, the QSA effect will be the same multiplicative factor on $r_u$ and $r_s$. When calculating $\delta$, QSA will divide out. However if $\delta$' is calculated, QSA will not directly divide out. The difference between $\delta$, $\delta$', and $\delta$' with QSA is shown in Figure \ref{fig:qsadeltafac}.

\section{Conclusions}
Previous assessments of the QSA effect have relied on measured isotope ratios \cite[e.g.][]{slodzian2004qsa}. These measurements can be affected by instrumental fractionation effects that can mask the QSA effect. To understand QSA better, it is necesssary to directly study the arrival of ions at the EM. In this paper, we have reassessed the QSA effect using measured pulse-height and arrival-time distributions. These measurements have allowed us to estimate the distribution of multiply emitted ions. We found that multiply emitted ions are overdispersed compared to a Poisson
distribution, making the traditional Poisson model of QSA, and its correction, invalid.  The combined effects of the secondary ion energy distribution, the secondary ion flight length, the secondary ion acceleration voltage, and the electron multiplier deadtime make it impractical to correct QSA by either an analytical or semi-empirical methodology. The best approach is to normalize unknown measurements with the appropriate standard, and calculate $\delta$ values appropriately.

\section{Acknowledgements} We thank Clive Jones (Washington University in St.\ Louis) for many invigorating discussions. This work was supported by NASA grant NNX14AF22G to RCO.

\section*{References}

\bibliography{mybibfile}

\end{document}